\documentclass{aa}
\usepackage{graphics}
\usepackage{graphicx}
\usepackage{txfonts}
\usepackage{afterpage}

\def\bc{\begin{center}}
\def\ec{\end{center}}
\def\be{\begin{equation}}
\def\ee{\end{equation}}
\def\bea{\begin{eqnarray}}
\def\eea{\end{eqnarray}}

\def\mkm{\mu{\rm m}}

\begin{document}
\title{From interstellar abundances to grain composition:
the major dust constituents Mg, \protect{Si,} and Fe}

\author{ N.V.~Voshchinnikov\inst{1,2,3}
         and
         Th.~Henning\inst{1}
         }
\authorrunning{Voshchinnikov and Henning}
\titlerunning{From interstellar abundances to grain composition}

\institute{
Max-Planck-Institut f\"ur Astronomie, K\"onigstuhl 17, D-69117 Heidelberg, Germany,
e-mail: {\tt voshchinnikov@mpia.de}
\and
Sobolev Astronomical Institute,
St.~Petersburg University, Universitetskii prosp. 28,
           St.~Petersburg, 198504 Russia,
e-mail: {\tt nvv@astro.spbu.ru}
\and
 Isaac Newton Institute of Chile, St.~Petersburg Branch
}
\date{Received ..., 2009 / accepted ..., 2010}

  \abstract{
We analyse observational correlations for
three elements entering into the composition of
interstellar silicate and oxide grains.
Using current solar abundances,
we converted the gas phase abundances into dust phase abundances
for 196 sightlines.
We deduce a sharp difference in abundances for sightlines located at
low ($|b|<30\degr$) and high ($|b|>30\degr$) galactic latitudes.
For {\rm high-latitude} stars, the ratios Mg/Si and Fe/Si in dust are close
to 1.5. For disk stars they are reduced to ${\rm Mg/Si} \sim 1.2$ and
${\rm Fe/Si} \sim 1.05$.
The derived numbers indicate that  1) the dust grains  cannot be the mixture of
silicates with olivine and pyroxene composition only,
and some amount of magnesium or iron (or both) should be in
another population and that 2) the destruction of Mg-rich grains
in the warm medium is more {\rm effective} than for Fe-rich grains.
We reveal a decrease in dust phase abundances and
correspondingly an increase in gas phase abundances with distance $D$ for
stars with $D\ga 400$\,pc. We attribute this to
{\rm an observational selection effect:}
a systematic trend toward lower  observed hydrogen
{\rm column} density for
distant stars. We find differences in abundances
for disk stars with low ($E({\rm B-V}) \la 0.2$) and
high ($E({\rm B-V}) \ga 0.2$) reddenings that reflect the distinction
between the sightlines passing through diffuse and translucent
interstellar clouds. For Scorpius--Ophiuchus,  we
detect a {\rm uniform}  increase in dust phase abundances of Mg and Si with
an increase in the ratio of total to selective extinction  $R_{\rm V}$
and a decrease in {\rm the strength of the far-UV extinction.}
This is the first evidence of growth of Mg-Si grains
{\rm due to accretion} in the interstellar medium.

      }
\maketitle

\section{Introduction}

Interstellar space is filled with gas and dust.
Both components interact with each other.
Atoms and molecules collide with solid particles and
cause grain growth or destruction (sputtering).
It depends on the relative velocity.
An examination of these processes is based on analysis of
the observed gas phase abundances.
Spectroscopic studies of interstellar UV absorption lines started
in the 1970s have revealed a deficit of heavy elements in the ISM in
comparison with cosmic ({\rm solar reference}) abundances
(Spitzer \& Jenkins,~\cite{sj75}). The missing atoms were assumed
to be tied up in solid particles
that opened an indirect way to investigate the element
composition of interstellar dust.
Modelling of interstellar extinction
(e.g., Li \& Greenberg,~\cite{lg97}; Zubko et al.~\cite{zda04};
Voshchinnikov et al.~\cite{vihd06}) has demonstrated that cosmic abundance
constraints might be crucial in deciding on modern dust models.

Cosmic abundances of heavy elements obtained from
spectroscopic studies of ordinary stars
(Snow \& Witt,~\cite{sw96}; Przybilla et al.~\cite{prz08})
and a decrease in the estimates of metal abundances in
the solar atmosphere  over the past years
(Asplund et al.~\cite{ags04, agss09}) essentially
limited the number of atoms incorporated into dust particles.
In this situation, accurate determination and analysis of gas phase
abundances are especially important.

A quantitative theory of element depletions is lacking.
First phenomenological  models showed a possible dependence of
depletions on the element equilibrium condensation temperature
(Field,~\cite{f74}) and the first or second element ionization potential
(Snow,~\cite{snow73}; Tabak,~\cite{tab79}).
The dependence of gas phase
abundances on hydrogen column density,
fraction of molecular hydrogen, distance, location, etc.,
were also considered
(Tarafdar et al.~\cite{tph83}; Harris et al.~\cite{hgb84};
Jenkins et al.~\cite{jss86}; see also references in
Jensen,~\cite{jens07} and Jenkins,~\cite{j09}).

A fresh approach to the problem of gas phase abundances has been devised by
Jenkins~(\cite{j04,j09}; hereafter J09) who investigated general patterns in
depletions of 17 elements. He finds that the propensity of
an element X to convert from gas to solid phase can be described
by a linear equation with coefficient $A_X$. The values of $A_X$
are assumed to be the same for all sightlines, while the individual
sightlines can be characterized by a depletion factor $F_*$ which
is common to all elements. This means that all elements are depleted
in unison independently of local physical conditions.

\begin{figure*}
\bc
\resizebox{\hsize}{!}{\includegraphics{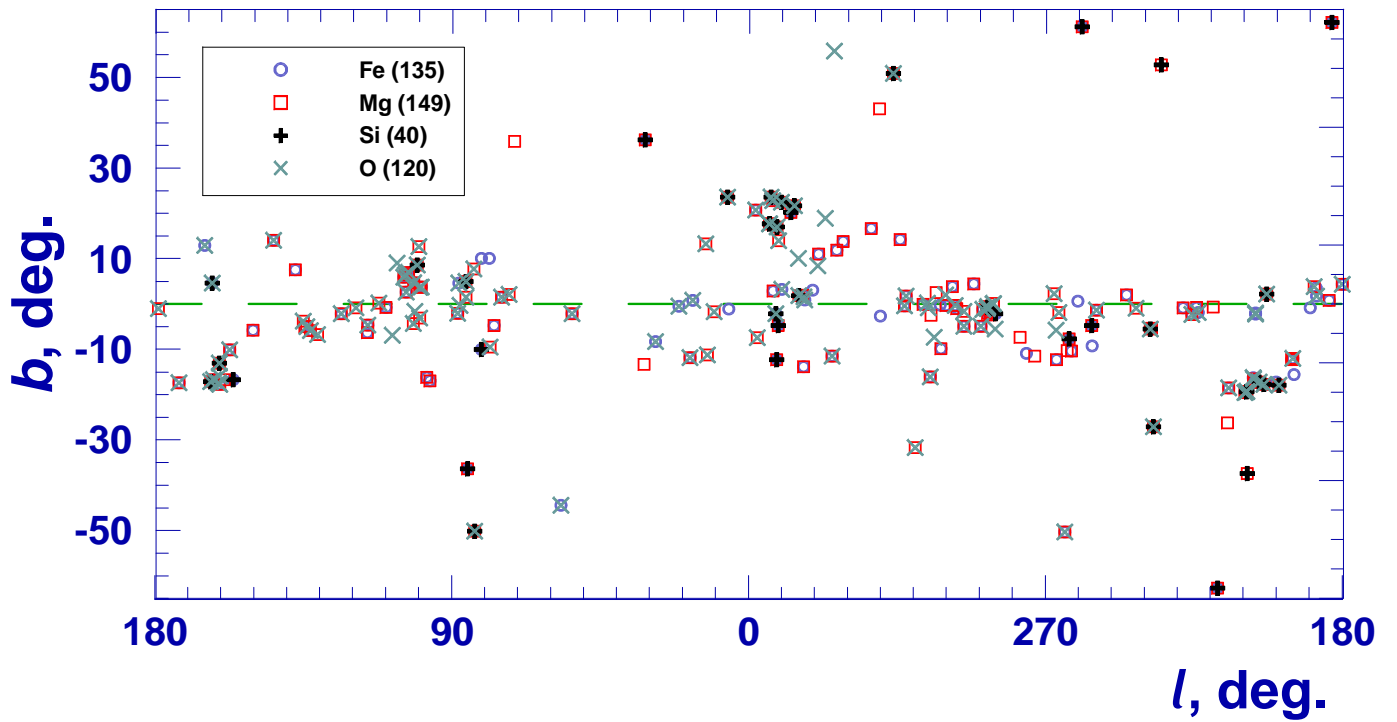}}
\caption{Galactic distribution of stars studied in this paper.
Sightlines with measured Fe, Mg, Si, and O are shown by different
symbols. Number of stars considered is indicated in parentheses
in the legend.}
\label{all-lb}\ec
\end{figure*}

In this paper, we exploit a more traditional approach by
trying to keep individual features of separate sightlines.
We investigate  correlations in abundances of
three elements Mg, Si, and Fe, which are classified as {\it major}
dust constituents (Jones,~\cite{jones99}).
These elements along with the {\it primary} element O can be
incorporated in solid phase in the form of Mg-Fe silicates,
metal particles, or oxides.
We  outline oxygen abundances,
which will be fully discussed elsewhere.
Another primary element, C, cannot be studied in such detail
since the number of sightlines with measured carbon abundances does not
exceed 20 (Sofia et al.~\cite{slmc04}; J09),
and at least in six directions  they have  been
revised downward in comparison with earlier estimates
(Sofia \& Parvahti,~\cite{sp09}).
Our main goal is to establish what could be the real dust phase
abundances and whether they could rule out ambiguity in
modelling interstellar extinction, polarization,
and spectral IR features.

\section{Sample of stars and first analysis}

\subsection{Definitions}

The abundance of an element in the interstellar medium is determined
as a number of atoms relative to that of hydrogen, [X/H], where X
(or $N({\rm X})$) and H (or $N({\rm H})=N({\rm HI})+2N({\rm H_2})$) are
the column densities of an element X and hydrogen in a given direction.
The abundances by number are often expressed
as the number of X atoms per $10^6$ hydrogen nuclei (parts per million,
ppm, hereafter).

Usually, the gas phase abundances of most elements
[X/H]$_{\rm g}$ are smaller
than the corresponding `cosmic' (reference, solar) abundances.
The depletion\index{Depletion of elements} of an element X is defined by
\begin{equation}
D_{\rm X} = \left. \left[\frac{\rm X}{\rm H}\right]_{\rm g} \right/
            \left[\frac{\rm X}{\rm H}\right]_{\rm cosmic}. \label{dx}
\end{equation}
The logarithmic quantities
\begin{equation}
\delta_{\rm X} = \log D_{\rm X} = \log \left[\frac{\rm X}{\rm H}\right]_{\rm g} -
         \log \left[\frac{\rm X}{\rm H}\right]_{\rm cosmic} \label{dx1}
\end{equation}
are also used\footnote{The bracketed notation [X/H] and the
units {\it ppm} are traditionally utilized when the dust phase abundances are studied
(e.g., Greenberg,~\cite{g78}; Mathis,~\cite{m96}; Voshchinnikov,~\cite{v04}).}.

\subsection{Data}

We assume that interstellar atoms are
in the dominant ionization stage for HI regions: OI, MgII, SiII, and FeII.
The contribution of neutral atoms to the total column density of magnesium,
silicon, and iron can be neglected
(see, e.g., data of Savage \& Bohlin,~\cite{sb79} and
Gnaci\'nski \&  Krogulec,~\cite{gnas06}).
The presence of the HII regions on the line of sight may lead to
some fraction of atoms in a stage of ionization above the
preferred one. To exclude this effect, J09 deduces his fits
for the sightlines with $N({\rm H}) > 3 \times 10^{19}$\,cm$^{-2}$,
although later he finds some departure from a linear trend between
the depletion factors $F_*$ and the logarithm of the average density for
sightlines with {\rm hydrogen column} densities lower than $3 \times 10^{19}$\,cm$^{-2}$.

\begin{figure}
\bc
\resizebox{\hsize}{!}{\includegraphics{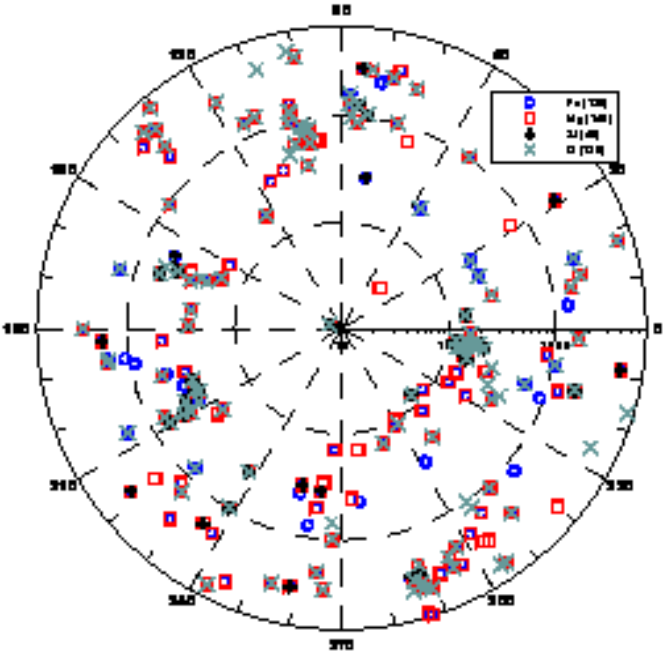}}
\caption{Distance distribution of stars studied in this paper in polar
representation. The Galactic centre is to the right.
The distance from the Sun is plotted in logarithmic scale.
Sightlines with measured Fe, Mg, Si, and O are shown with different
symbols. Number of stars is indicated in parentheses.}
\label{all-pol}\ec
\end{figure}

Our list of stars includes different targets with
measured gas phase abundances of oxygen, magnesium, silicon, or iron.
We transform all data into the unified (standard) system of oscillator
strengths as given in Table~1 of J09. We also use the hydrogen
column density from J09 when available\footnote{$N$({\rm H})
for star CPD\,--59\,2603 is taken from Jensen \& Snow (\cite{jens07a}).}.
The final sample contains 196 sightlines with 1$\sigma$ errors
{\rm (see Table~\ref{t00} in Appendix).}
Observational data are taken from J09,
Cartledge et al.~(\cite{cart04}, \cite{cart08}), and
Jensen et al.~(\cite{jens05})  for oxygen (120 sightlines);
J09, Jensen \& Snow~(\cite{jens07b}), Cartledge et al.~(\cite{cart06}),
Howk et al.~(\cite{howk99}),
and Gnaci\'nski \&  Krogulec~(\cite{gnas06}) for magnesium (149);
Jensen~(\cite{jens07}),  Gnaci\'nski \&  Krogulec~(\cite{gnas06}), and J09
for silicon (40); and J09,  Jensen \& Snow~(\cite{jens07a}),
Snow et al.~(\cite{srf02}), and Miller et al.~(\cite{mil07})
for iron (135). For two sightlines (HD~93521 and HD~215733), where
the data for separate velocity components are obtained, we take the total
column densities.
The galactic distribution of all 196 stars is
plotted in Fig.~\ref{all-lb}. Overlapping symbols indicate that,
for a given sightline  measurements were made for more than one element.
Figure~\ref{all-pol} illustrates the distance distribution of stars in
the projection on the galactic plane.

The abundances were supplemented with stellar distances $D$,
colour excesses $E({\rm B-V})$ and characteristics
of the extinction curves: the ratio of total to selective
extinction  $R_{\rm V}=A_{\rm V}/E({\rm B-V})$ and the parameters of
UV extinction as suggested by Fitzpatrick \& Massa~(\cite{fm90}, \cite{fm07}).
Extinction data are  collected from papers of
Fitzpatrick \& Massa~(\cite{fm07}), Valencic et al.~(\cite{vcg04}),
Wegner~(\cite{ww02}, \cite{ww03}), Patriarchi et al.~(\cite{pmp03})
and the papers with published abundances cited above.

We recalculated all colour excesses $E({\rm B-V})$ published by J09
using normal colours of stars from Strai\u{z}ys~(\cite{Str92}) and Landolt-B\"ornstein
and spectral types from Bowen et al.~(\cite{bow08}) and J09.
This permits avoiding the difficulties with the negative values of
$E({\rm B-V})$ obtained by J09. Using the 2MASS K-magnitudes from the Simbad
database, correction to the Johnson
system given by Bowen et al.~(\cite{bow08}) and normal
colours $({\rm V-K})_0$
from Strai\u{z}ys~(\cite{Str92}) and Winkler~(\cite{wink97}), we estimated
colour excesses $E({\rm V-K})$ for several stars. Next we found the ratio
$R_{\rm V}$ with the aid of the relation
$R_{\rm V}= 1.1\, E({\rm V-K})/E({\rm B-V})$ (Voshchinnikov \&  Il'in, \cite{vi87}).

\begin{figure}
\bc
\resizebox{8.0cm}{!}{\includegraphics{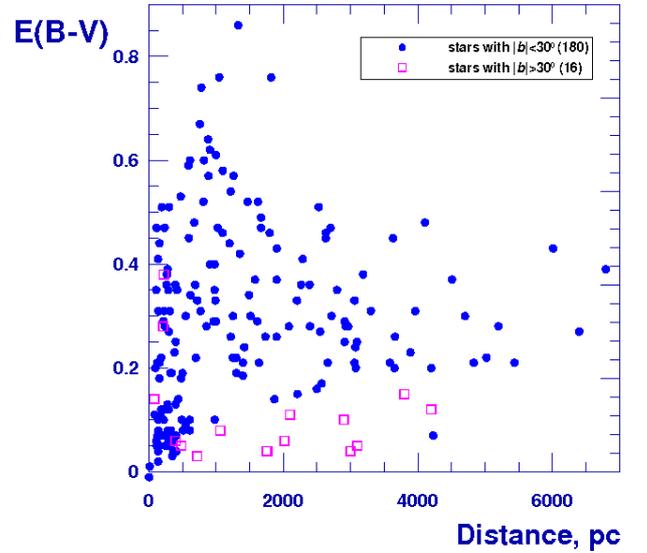}}
\caption{Distance distribution of colour excess $E({\rm B-V})$ for
stars studied in this paper.
Open squares show stars located at galactic latitudes $|b|>30\degr$.
}
\label{ebv-d}\ec
\end{figure}

\begin{figure}[htb]
\bc
\resizebox{\hsize}{!}{\includegraphics{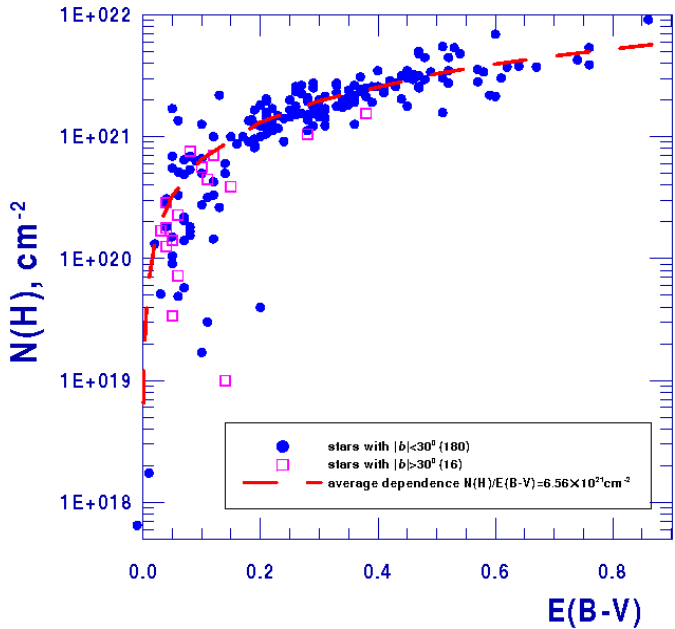}}
\caption{Total hydrogen column density as function of the colour
excess for stars studied in this paper. The dashed line shows the
average dependence between $N({\rm H})$ and $E({\rm B-V})$ deduced
by Ryter~(\cite{ryt96}).}
\label{nh}\ec
\end{figure}


From the distance distribution of colour excesses 
plotted in Fig.~\ref{ebv-d}, it follows that
the major part of stars have $E({\rm B-V}) \la 0.6$.
This means that  they are located behind diffuse atomic and molecular clouds or
translucent interstellar clouds (see Table~1 in Snow \& McCall,~\cite{sn06}
for classification).
In Fig.~\ref{ebv-d} we separate 16 stars
located at high galactic latitudes $|b|>30\degr$\footnote{{\rm We include
HD~38666 ($b=-27.1\degr$) where anomalous high gas phase
abundances of Mg, Si, and Fe are observed in this list of stars.}}.
Previous considerations (e.g., Savage \& Sembach,~\cite{sav96a}) indicate
that the gas has smaller depletions in the lower halo
in comparison with gas located in the galactic disk.
The total hydrogen column density is smaller
for stars observed at high galactic latitudes (Fig.~\ref{nh}).
Figure~\ref{nh} allows one to study the gas-to-dust ratio in the direction
of stars from our sample.
Ryter~(\cite{ryt96}) deduce the average gas-to-dust ratio
\be
\frac{N({\rm H})}{E{\rm (B-V)}} = 6.56 \times 10^{21} \,\,
{\rm atoms \, cm^{-2} \, mag^{-1}}\,,
\label{ryter}\ee
where the contribution of ionized hydrogen is neglected.
From Fig.~\ref{nh} it can be seen that the dependence  of $N({\rm H})$
on $E({\rm B-V})$ does not strongly deviate from the average dependence
if $E({\rm B-V}) \ga 0.2$. Near this value there is the border between
diffuse and translucent interstellar clouds (Snow \& McCall,~\cite{sn06}).
Observational errors are larger if the sightline crosses a diffuse cloud.
For stars seen through the translucent clouds,
the ratio $N({\rm H})/E{\rm (B-V)}$ lies within rather narrow limits from
$\sim 3 \times 10^{21}\, {\rm cm^{-2}\, mag^{-1}}$ to
$\sim 1 \times 10^{22}\, {\rm cm^{-2}\, mag^{-1}}$.
For almost all sightlines considered by us, the criterion of J09
($N({\rm H}) > 3 \times 10^{19}$\,cm$^{-2}$) is satisfied.
Two points in the lower left corner of Fig.~\ref{nh}
correspond to HD~34029 (Capella)\footnote{J09 gives
the observed colour ${\rm B-V}$ not $E({\rm B-V})$ for Capella in Table~2.}
and HD~48915 (Sirius).
The colour excesses for them are very small and uncertain, therefore
we exclude these two stars from further analysis.


\begin{figure}[!htb]
\bc
\resizebox{8.0cm}{!}{\includegraphics[bb=75 304 511 676,clip]{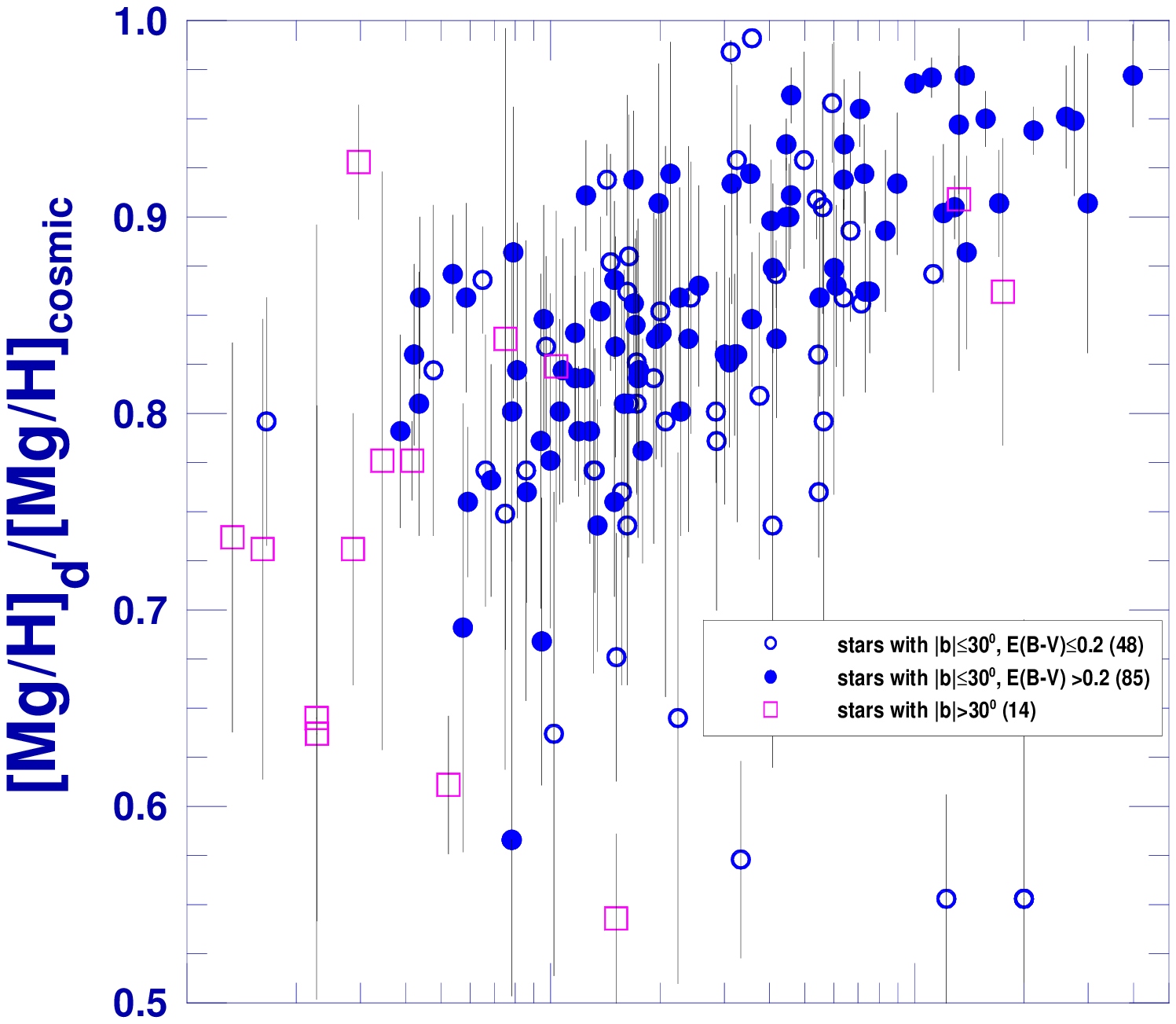}}
\resizebox{8.0cm}{!}{\includegraphics[bb=75 304 511 676,clip]{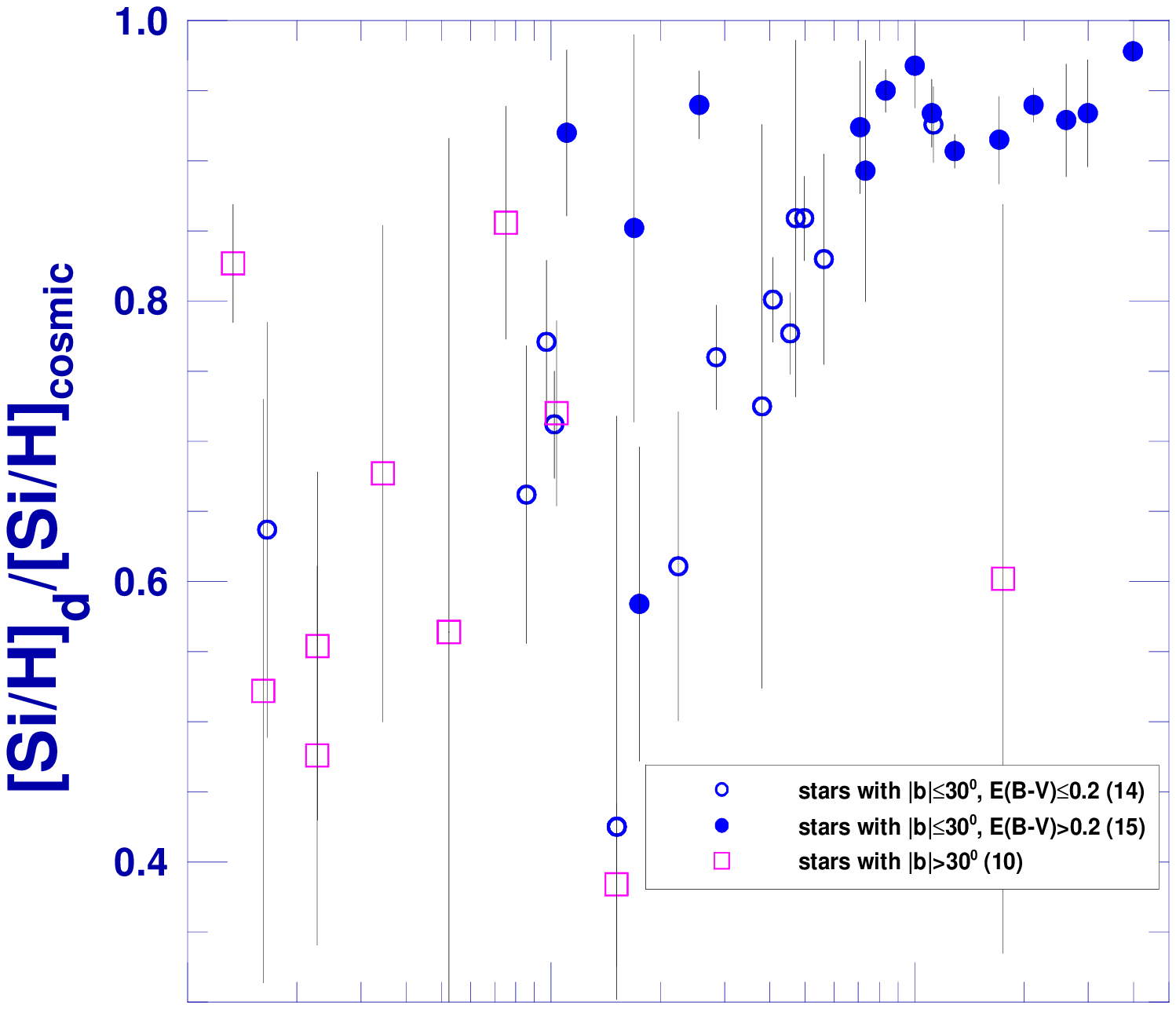}}
\resizebox{8.0cm}{!}{\includegraphics[bb=75 247 511 676,clip]{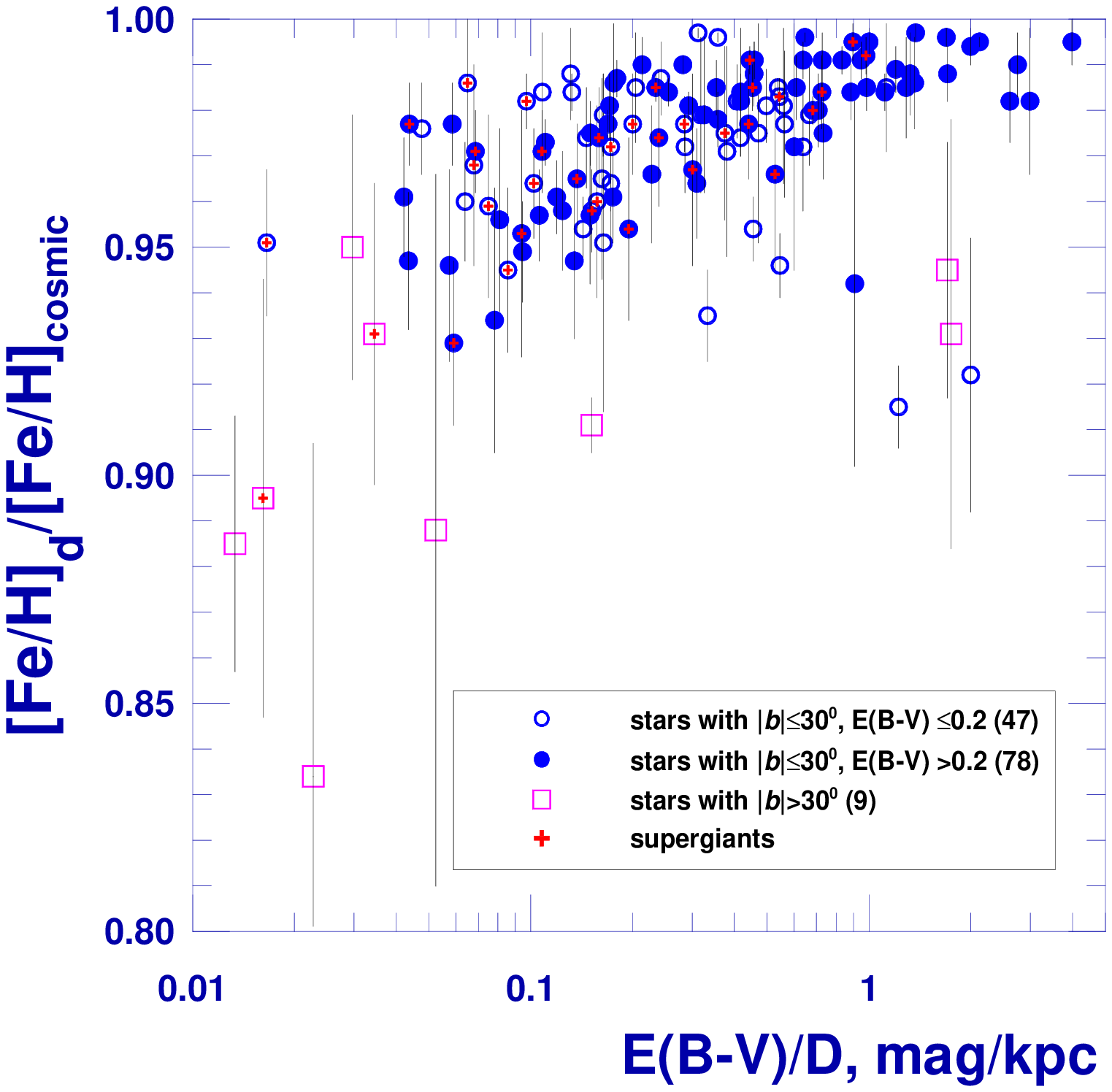}}
\caption{Relative dust phase abundances of Mg, Si, and Fe with 1$\sigma$
error bars in dependence on average reddening $E({\rm B-V})/D$.
Circles and squares show data for stars with $|b|\leq 30\degr$ and
$|b| > 30\degr$, respectively.
The number of stars is indicated in parentheses.
Crosses in lower panel show data for supergiants.}
\label{mgd}\ec
\end{figure}

\afterpage{\clearpage}

\subsection{Reference abundances}

We use, as a starting point,
the solar abundances from Asplund et al.~(\cite{agss09})
as cosmic abundances. These data are justified
and are rather close to the modern solar
system abundances recommended by Lodders et al.~(\cite{lpg09}).
Taking
[O/H]$_{\sun}= 490$~ppm, [Mg/H]$_{\sun}= 39.8$~ppm,
[Si/H]$_{\sun}= 32.4$~ppm, and [Fe/H]$_{\sun}= 31.6$~ppm,
and neglecting the errors in the reference abundances,
we converted the gas phase abundances into the dust phase abundances
\bea
\left[\frac{\rm X}{\rm H}\right]_{\rm d} =
        \left[\frac{\rm X}{\rm H}\right]_{\rm cosmic} -
        \left[\frac{\rm X}{\rm H}\right]_{\rm g}
=   \left[\frac{\rm X}{\rm H}\right]_{\rm cosmic} \left(1-D_{\rm X}\right)
 \nonumber \\
= \left[\frac{\rm X}{\rm H}\right]_{\rm cosmic} \left(1-10^{\delta_{\rm X}}\right).
\label{ddust}
\eea

\subsection{Choice of external parameter and selection effects}\label{sel}

Previous considerations reveal  correlations between
element depletion and an `external' parameter characterizing the gas
density in the line of sight: $N({\rm H})$,
$\langle n({\rm H}) \rangle =N({\rm H})/D$,
$f({\rm H}_2)= 2N({\rm H_2})/N({\rm H})$
(see Jensen,~\cite{jens07} and J09 for detailed discussion).
Clear trends toward increasing gas depletion (and correspondingly
growth of dust phase abundances) are found for Mg, Si, and Fe
with respect to $\langle n({\rm H}) \rangle$.
Sometimes, variations in depletion with distance are also considered
(Cartledge et al.~\cite{cart06}, \cite{cart08}; Jensen,~\cite{jens07}).

In deciding on the parameter connecting dust grains and
abundances, we are restricted by reddening $E({\rm B-V})$,
extinction $A_{\rm V}$ and average reddening $E({\rm B-V})/D$ or
extinction $A_{\rm V}/D$.
In order to determine extinction, we must know the ratio $R_{\rm V}$,
which is not  easily  detected from observations
(see discussion in Strai\u{z}ys,~\cite{Str92}).
It is also important that we only know total (summary) extinction or
reddening and cannot separate individual clouds on the line
of sight.


For further analysis, we choose the average reddening $E({\rm B-V})/D$
as an external parameter because $E({\rm B-V})$ and $D$ are usually
well known for different sightlines. Colour excess $E({\rm B-V})$ (or reddening)
characterizes the amount of dust on the line of sight and the properties
of dust particles. It  can be calculated as
$$
E({\rm B-V}) = A_{\rm B} - A_{\rm V}
\approx 1.086 (\langle C_{\rm ext, B} \rangle -
\langle C_{\rm ext, V} \rangle ) N_{\rm d}\,,
$$
where $N_{\rm d}$ is the dust column density and
$\langle C_{\rm ext, B} \rangle$, $\langle C_{\rm ext, V} \rangle$ are
average extinction cross sections.
Sightlines with low reddening may be the result of the absence of dust
(low $N_{\rm d}$-value) or similar cross sections
in the B and V bands. The latter can be interpreted as the presence
of large grains producing neutral extinction.
Perhaps, such sightlines have higher hydrogen column density
than average and appear as points above the curve
at the left upper part of Fig.~\ref{nh}. However, one should keep in mind that
the error of the colour excess is larger for lower values
of $E({\rm B-V})$\footnote{The polarimetric data obtained for stars
with low reddening give the polarization efficiency close to average one
$P/E({\rm B-V}) = 3 \%/{\rm mag}$
(Berdyugin et al.~\cite{btp10}).}.


Relative dust phase abundances of Mg, Si, and Fe are plotted in Fig.~\ref{mgd}
in the form [X/H]$_{\rm d}$/[X/H]$_{\rm cosmic} = 1-D_X$ as a function
of $E({\rm B-V})/D$.
Using Eq.~(\ref{ryter}), the average reddening can be translated into
average gas density:
$$
\langle n({\rm H}) \rangle [{\rm cm}^{-3}] \approx
2.13\,\times \, E({\rm B-V})/D [{\rm mag/kpc}].
$$
Stars in our sample are divided into two groups depending on
their position in the Galaxy: stars {\rm at high galactic latitudes}
($|b|>30\degr$) and stars located near the disk
($|b| \leq 30\degr$), which in turn are separated as
sightlines passing through diffuse ($E({\rm B-V}) \leq 0.20$) and
translucent ($E({\rm B-V}) > 0.20$) clouds.
It can be seen that the abundances for {\rm high-latitude}
and disk sightlines are quite different.
For {\rm high-latitude} stars, the element fraction in dust is lower
and does not depend on the average reddening.
This is well established and was interpreted in the homework
of the  Spitzer~(\cite{sp85}) model of the different grain composition
in the warm  and cold phases of the ISM.

For disk stars, independent of reddening,
there is a smooth decrease in the dust phase abundances of Mg, Si,
and Fe when $E({\rm B-V})/D$ grows smaller.
This is expected in the context of previous findings
(Jensen,~\cite{jens07}, J09).
The reason for such behaviour may be some physical processes
{lower cosmic abundances outside the solar environment}
or observational selection. Selection effects can appear
for distant stars as they may be more luminous and produce
more UV flux that can modify gas abundances.
To check this we mark out in Fig.~\ref{mgd} (lower panel)
19 supergiants. It is evident that the directions to supergiants
are not peculiar and fall well into the general pattern of
disk and {\rm high-latitude}
stars. Therefore, differences in the types of material probed
along more  and less luminous stars seem cannot
cause the trends observed in Fig.~\ref{mgd}.

\begin{figure}
\bc\resizebox{\hsize}{!}{\includegraphics{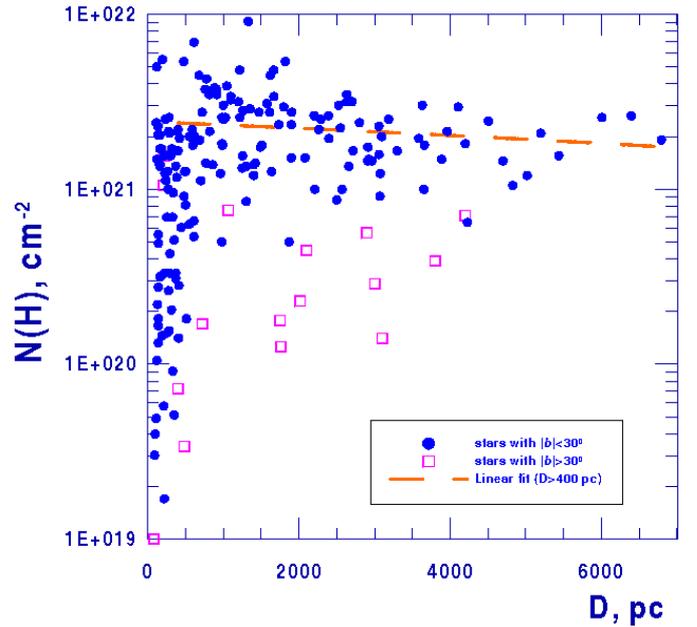}}
\caption{Distance dependence of total hydrogen column density
for stars studied in this paper. The dashed line is the fit
for stars with $D>400$\,pc.}
\label{nhd}\ec
\end{figure}

\begin{figure}[htb]
\bc\resizebox{\hsize}{!}{\includegraphics{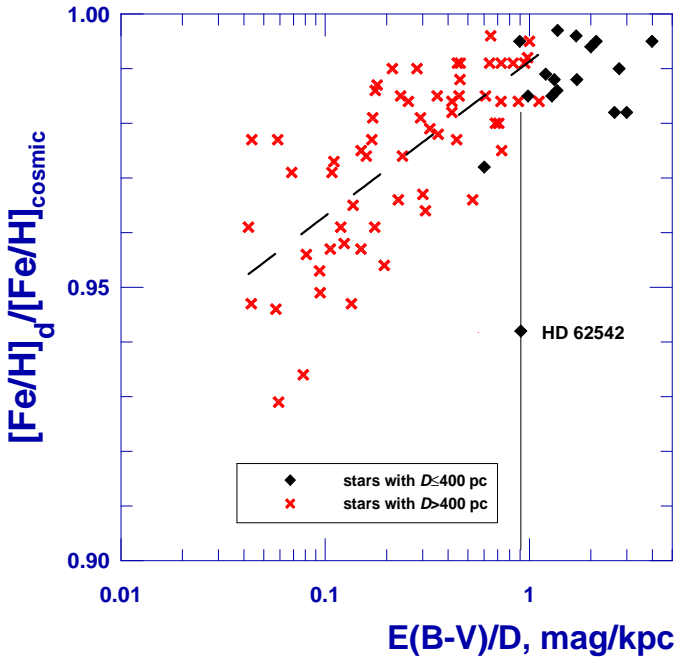}}
\caption{Relative dust phase abundances of  Fe
as a function  of average reddening $E({\rm B-V})/D$ for
stars with $|b|\leq 30\degr$ and $E({\rm B-V})>0.20$.
Rhombuses and crosses show data for stars with
$D \leq 400$\,pc and $D>400$\,pc, respectively.
The dashed line is the fit.}
\label{fe2}\ec
\end{figure}

The true reason of decreasing dust phase abundances
(and correspondingly increasing  gas phase
abundances) for smaller $E({\rm B-V})/D$ and
$\langle n({\rm H}) \rangle$ is evident from Fig.~\ref{nhd}
{\rm (see also Fig.~\ref{ebv-d}).}
{\it A systematic tendency exists for a decrease in the observed hydrogen
{\rm column} density {\rm and colour excess}
for distant stars.} In our sample (178 sightlines
with $|b|\leq 30\degr$), the decrease is observed for stars with
$D\ga 400$\,pc {\rm with
the gas-to-dust ratio remaining almost constant for clouds
located at different distances.}
As a result of lower hydrogen column density for distant stars,
we obtain a clear trend for abundances as a function of
{\rm any} parameter
related to the gas or dust density or distance.
Figure~\ref{fe2} illustrates this {\rm observational selection effect}.
It shows the ratio [Fe/H]$_{\rm d}$/[Fe/H]$_{\rm cosmic}$ in dependence on
$E({\rm B-V})/D$ for stars observed through translucent interstellar clouds.
Apparently, {\rm many} previous correlations of interstellar abundances
reflect a decrease in $N({\rm H})$ {\rm and $E({\rm B-V})$}
with increasing $D$.
It looks like we observe less and less dense clouds when the distance
grows. Evidently, this problem requires further investigation,
{\rm especially observations of high-density longer sightlines.}

\setcounter{table}{1}
\begin{table*}[Htb!]
\begin{center}
\caption{Mean value of element abundance in dust phase in ppm
with 1$\sigma$ error}
\begin{tabular}{lllllll} \hline
\noalign{\smallskip}
& \multicolumn{1}{c}{$\langle$[Mg/H]$_{\rm d} \rangle$} &
\multicolumn{1}{c}{$\displaystyle\frac{\langle[{\rm Mg/H}]_{\rm d} \rangle}{[{\rm Mg/H}]_{\rm cosmic}}$} &
\multicolumn{1}{c}{$\langle$[Si/H]$_{\rm d} \rangle$} &
\multicolumn{1}{c}{$\displaystyle\frac{\langle[{\rm Si/H}]_{\rm d} \rangle}{[{\rm Si/H}]_{\rm cosmic}}$} &
\multicolumn{1}{c}{$\langle$[Fe/H]$_{\rm d} \rangle$} &
\multicolumn{1}{c}{$\displaystyle\frac{\langle[{\rm Fe/H}]_{\rm d} \rangle}{[{\rm Fe/H}]_{\rm cosmic}}$} \\
\noalign{\smallskip}
\hline
\noalign{\smallskip}
all sightlines &  33.13 $\pm$ 2.35  & 0.832 $\pm$ 0.059 (147)$^*$&
                  25.01 $\pm$ 2.91  & 0.772 $\pm$ 0.090 (39)     &
                  30.64 $\pm$ 0.41  & 0.970 $\pm$ 0.013 (134)    \\
stars with $|b|\leq 30\degr$ \\
~~~$E({\rm B-V}) \leq 0.20$ & 32.31 $\pm$ 2.60  & 0.812 $\pm$ 0.065 (48)&
                                        23.97 $\pm$ 3.03  & 0.740 $\pm$ 0.093 (14)&
                                        30.65 $\pm$ 0.40  & 0.970 $\pm$ 0.013 (47)\\
~~~$E({\rm B-V}) > 0.20$ & 34.11 $\pm$ 1.96  & 0.857 $\pm$ 0.049 (85)&
                                     29.30 $\pm$ 1.48  & 0.904 $\pm$ 0.046 (15)&
                                     30.86 $\pm$ 0.31  & 0.977 $\pm$ 0.010 (78)\\
                   & {\it 36.90 $\pm$ 1.44}  & {\it 0.927 $\pm$ 0.036 (14)}&
                                     {\it 30.25 $\pm$ 0.76}  & {\it 0.934 $\pm$ 0.023 (6)}&
                                     {\it 31.16 $\pm$ 0.31}  & {\it 0.986 $\pm$ 0.010 (17)}\\
stars with $|b|> 30\degr$ & 29.98 $\pm$ 3.92  & 0.753 $\pm$ 0.099 (14)&
                 20.03 $\pm$ 4.90  & 0.618 $\pm$ 0.151 (10)&
                 28.68 $\pm$ 1.30  & 0.908 $\pm$ 0.041 (9) \\
\noalign{\smallskip}
\hline
\noalign{\smallskip}
\multicolumn{3}{l}{$^*$ Number of sightlines} \\
\multicolumn{5}{l}{Results for stars with $D\leq 400$\,pc and
$|b|\leq 30\degr$,  $E({\rm B-V}) > 0.20$ are given in italics.}
\label{tab1}
\end{tabular}
\end{center}
\end{table*}


\section{Results and discussion}
\subsection{Mean values and deviations}

Mean element abundances locked up in dust are given in Table~\ref{tab1}.
The data are presented for all sightlines
and separately for {\rm high-latitude}  and disk stars having low and high reddening,
respectively. Distinctions between the various groups are noticed.
For all three elements the following inequality is valid:
$$
\left \langle \left [{\rm {X}/{H}} \right ]_{\rm d}  \right\rangle_{|b|>30\degr} <
\left \langle \left [{\rm {X}/{H}} \right ]_{\rm d}  \right\rangle_{\rm |b|<30\degr,\,{\it E}({\rm B-V}) \leq 0.2} <
$$
$$
< \left \langle \left [{\rm {X}/{H}} \right ]_{\rm d}  \right\rangle_{\rm |b|<30\degr,\,{\it E}({\rm B-V}) > 0.2}.
$$
As follows from Fig.~\ref{mgd}, halo {\rm ($z \ga 150$\,pc)} and disk stars can
be easily distinguished by the average reddening.  For halo stars we have
$E({\rm B-V})/D \la 0.05$\,{\rm mag/kpc}. Exceptions are three short
{\rm high-latitude} sightlines
for HD~116658 ($b=+50.8\degr$, $D=80$\,pc), HD~203532 ($b=-31.7\degr$, $D=211$\,pc), and
HD~210121 ($b=-44.4\degr$, $D=223$\,pc). {\rm They
cross relatively dense clouds (see also Fig.~\ref{ebv-d}) and have
abundances similar to those of halo stars.}
The upper right corner of Fig.~\ref{mgd} is mainly filled by reddened
nearby stars. A dividing line between
stars with distances greater or less than 400\,pc passes near the value
$E({\rm B-V})/D \approx 1$\,{\rm mag/kpc} (Fig.~\ref{fe2}).
Because the data for stars with
$D \ga 400$\,pc seems to be `infected' with  observational bias,
we also calculated the mean abundances for nearby stars.
A noticeable growth of abundances occurs for reddened disk stars.
The results are shown in Table~\ref{tab1} in italics.

There are several sightlines where the dust phase abundances of Mg and Fe
are significantly lower than the general trends clearly seen in Fig.~\ref{mgd}.
The major part of `peculiar' sightlines is related to  stars with
$E({\rm B-V}) \leq 0.2$. Only two objects (HD~62542 and HD~99890)
are observed through translucent clouds, but the observational errors
are quite large in these cases. Note also that the shape of the UV
extinction curve in the direction of HD~62542 is very peculiar
(Voshchinnikov \& Das,~\cite{vd08}). It should be remembered
that almost all stars with anomalous low
dust phase abundances are located in Carina--Centaurus
at the galactic longitudes $l \approx 290\degr - 330\degr$.

We compared our results with data presented in Table~4 of J09, which
gives element depletion parameters corresponding to
the cases of `full depletion' ($F_* =1$) and `no depletion'
($F_* =0$). We transformed these parameters
into our reference system and find the relative dust phase abundances
of Mg, Si, and Fe. They are equal to
[Mg/H]$_{\rm d}$/[Mg/H]$_{\rm cosmic}$=0.94 and 0.44,
[Si/H]$_{\rm d}$/[Si/H]$_{\rm cosmic}$=0.94 and 0.25, and
[Fe/H]$_{\rm d}$/[Fe/H]$_{\rm cosmic}$=0.99 and 0.88
for the cases $F_* =1$ and $F_* =0$, respectively.
The data for the `full depletion' case are well
within our results (see Table~\ref{tab1} and Fig.~\ref{mgd}).
For another case (`no depletion'), the  abundances of J09
seem to be too low even for {\rm high-latitude} stars. This may be a result of
observational selection and smaller number of sightlines in comparison
with our sample.

\subsection{Correlations}

\enlargethispage{\baselineskip}

Using our data it is possible to plot the dust phase abundance of one
element against that of another element. Such diagrams clearly show
the existence of strong correlations between the abundances of Mg, Si,
and Fe for low reddened and distant stars (see, e.g.,
Cartledge et al.~\cite{cart06}; Miller et al.~\cite{mil07}).
However, these correlations trace the behaviour of
the hydrogen column density discussed in Sect.~\ref{sel}.

\begin{figure}[htb]
\bc
\resizebox{\hsize}{!}{\includegraphics{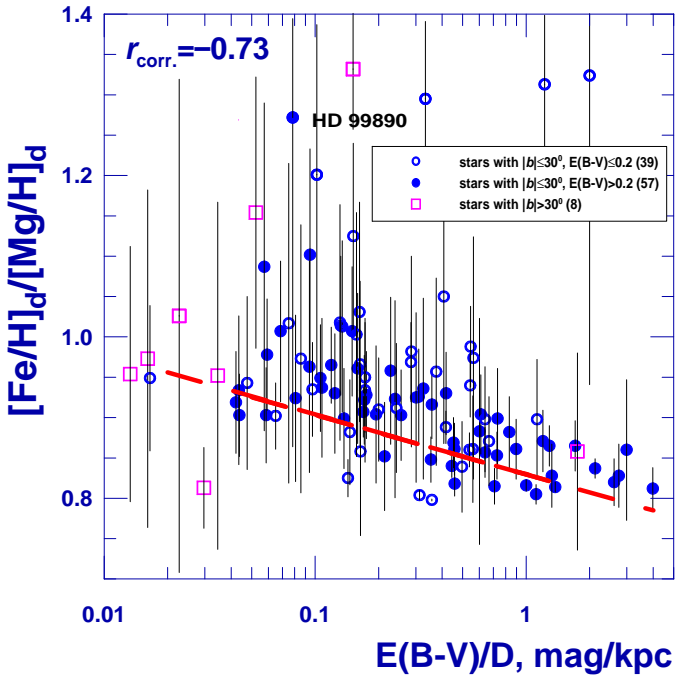}}
\caption{Ratio of dust phase abundances of Fe and Mg
in dependence on average reddening $E({\rm B-V})/D$.
Open and filled circles show data for disk stars with
$E({\rm B-V}) \leq 0.20$ and $E({\rm B-V}) > 0.20$, respectively.
Squares correspond to sightlines with $|b| > 30\degr$.
The bold dashed line is the linear regression fit for stars with $E({\rm B-V}) > 0.20$:
$[{\rm Fe}/{\rm H}]_{\rm d}/[{\rm Mg}/{\rm H}]_{\rm d}
=(-0.074 \pm 0.010) \log [E({\rm B-V})/D]
+ (0.830 \pm 0.003)$.}
\label{mg-fe}\ec
\end{figure}


To exclude the effect of $N_{\rm H}$ we plot the ratio
of the dust phase abundances of Fe to Mg in dependence on
the average reddening. The result is shown in Fig.~\ref{mg-fe} for
104 sightlines with the linear regression fit
for 56 disk sightlines with $E({\rm B-V}) > 0.20$ (HD~99890 was excluded)
as derived by  a $\chi^2$ minimization that
takes error bars into account.
The Pearson correlation coefficient for these sightlines is
$r_{\rm corr.}=-0.73$.
As follows from Fig.~\ref{mg-fe}, the amount of iron
grows slightly
in comparison with the amount of magnesium when
the average reddening decreases. Twelve stars
have distances $D<400$\,pc. They are located in the
bottom right part of Fig.~\ref{mg-fe} and have an almost constant
ratio ${\rm Fe/Mg} \approx 0.84$.

A similar behaviour can be observed if we compare
dust phase abundances of Mg or Fe and Si. However,
in this case the number of sightlines is three times less, because it
is dictated by the measurements of silicon (see Figs.~\ref{all-lb}
and \ref{all-pol}). For almost all
sightlines with measured Si we have measurements of Mg and Fe.
Therefore, one can consider the  composition of grains.

\subsection{Olivines,  pyroxenes, and... ?}\label{ol}

All elements considered by us are constituents of cosmic silicate
grains showing  a pronounced 9.7\,$\mu$m  feature observed in spectra of
a wide variety of objects (see Henning~\cite{he09}, for a recent review).
The origin of this feature is related to the stretching of the Si--O bond
in amorphous silicates with olivine (Mg$_{\rm 2x}$Fe$_{\rm 2-2x}$SiO$_4$) or
pyroxene (Mg$_{\rm y}$Fe$_{\rm 1-y}$SiO$_3$) stoichiometry,
where $0 \leq x, y \leq 1$.
If we assume that Mg, Si, and Fe are incorporated only
into Mg-Fe silicates, the ratio of (Fe+Mg)/Si in the dust phase must
be in the range from 1 to 2, while the ratios Mg/Si and Fe/Si
may vary from 0 to 2.


\begin{figure}[htb]
\bc
\resizebox{8.00cm}{!}{\includegraphics[bb=96 391 520 759,clip]{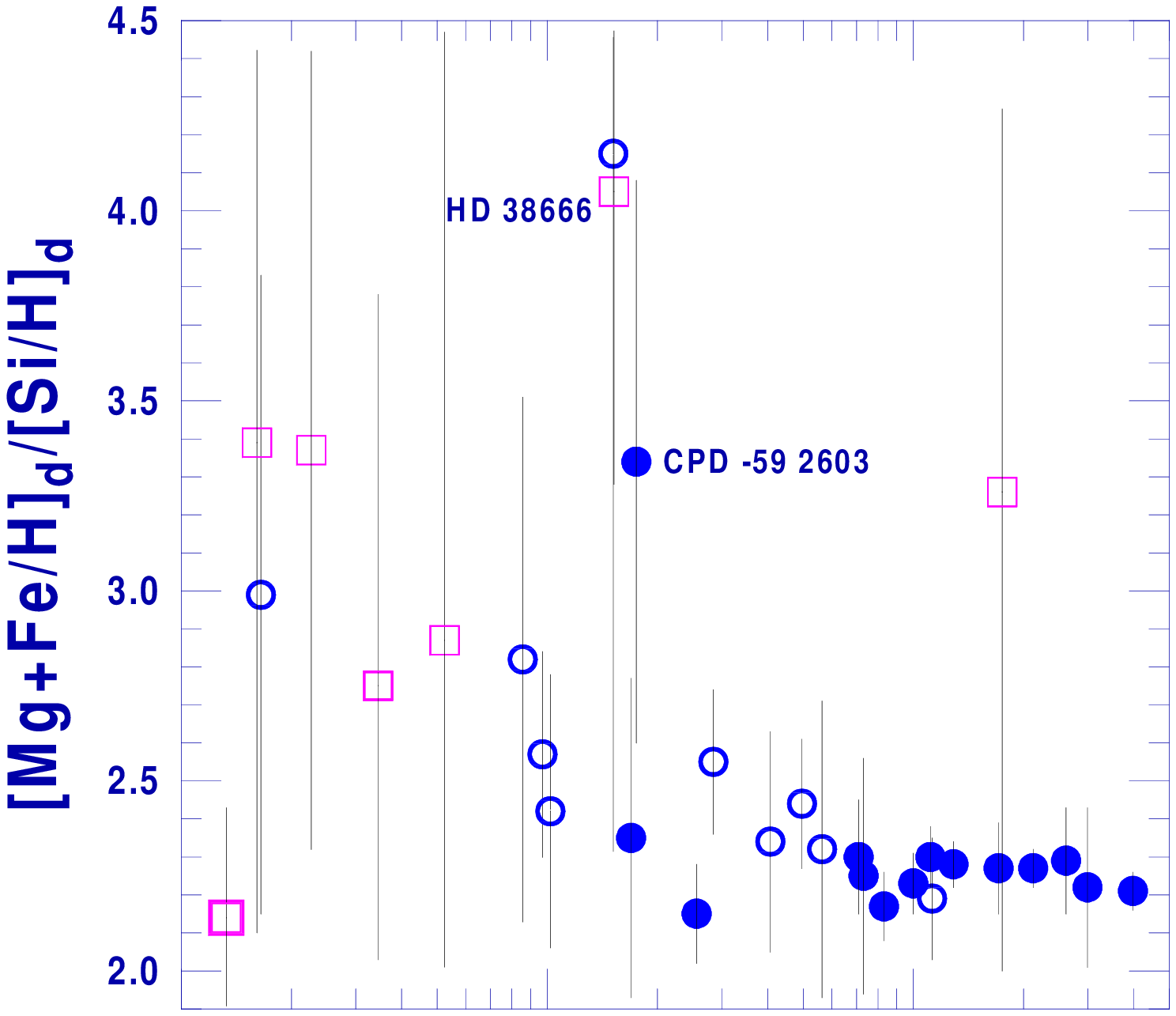}}
\resizebox{8.00cm}{!}{\includegraphics[bb=96 391 520 754,clip]{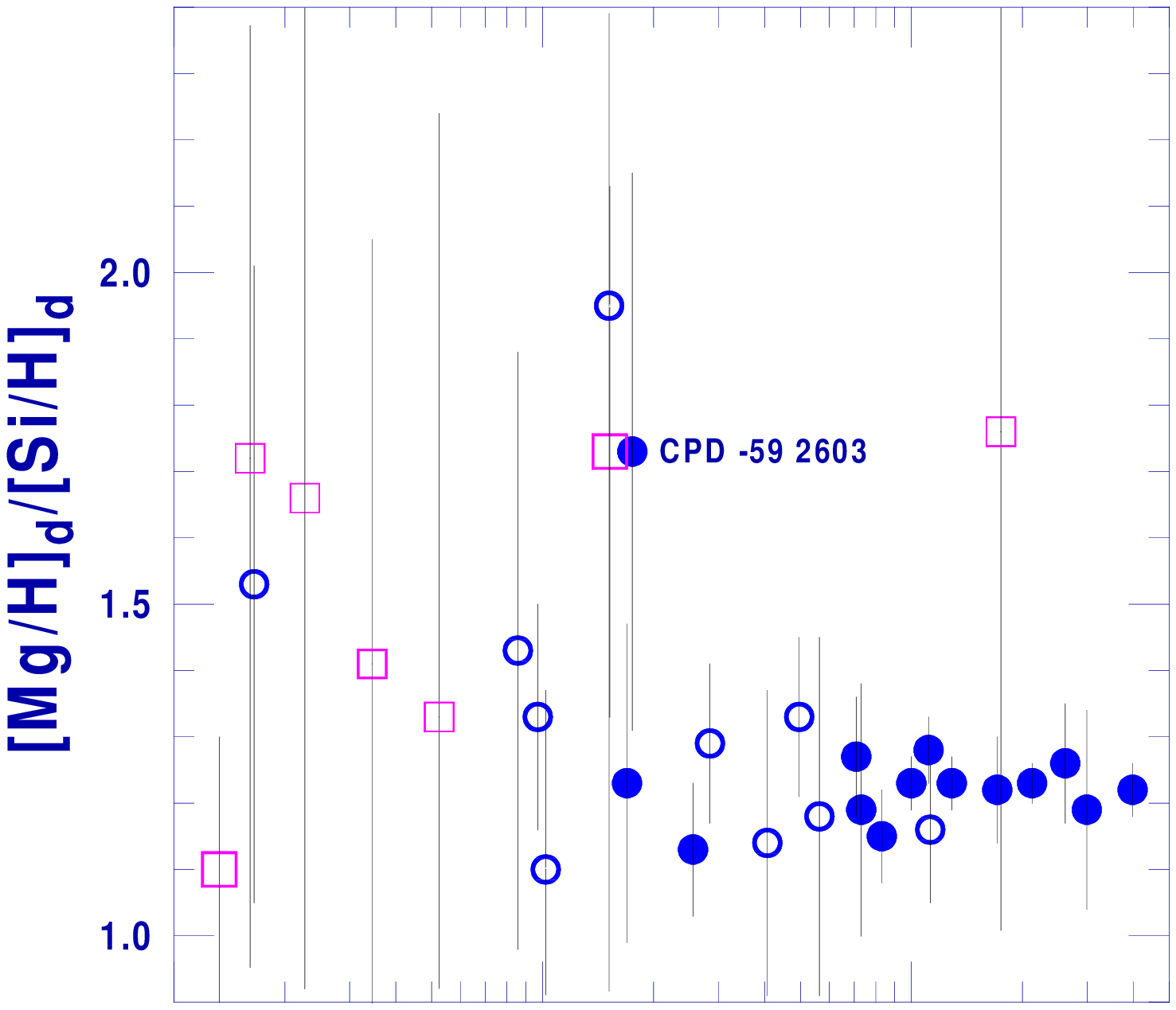}}
\resizebox{8.00cm}{!}{\includegraphics[bb=88 246 511 672,clip]{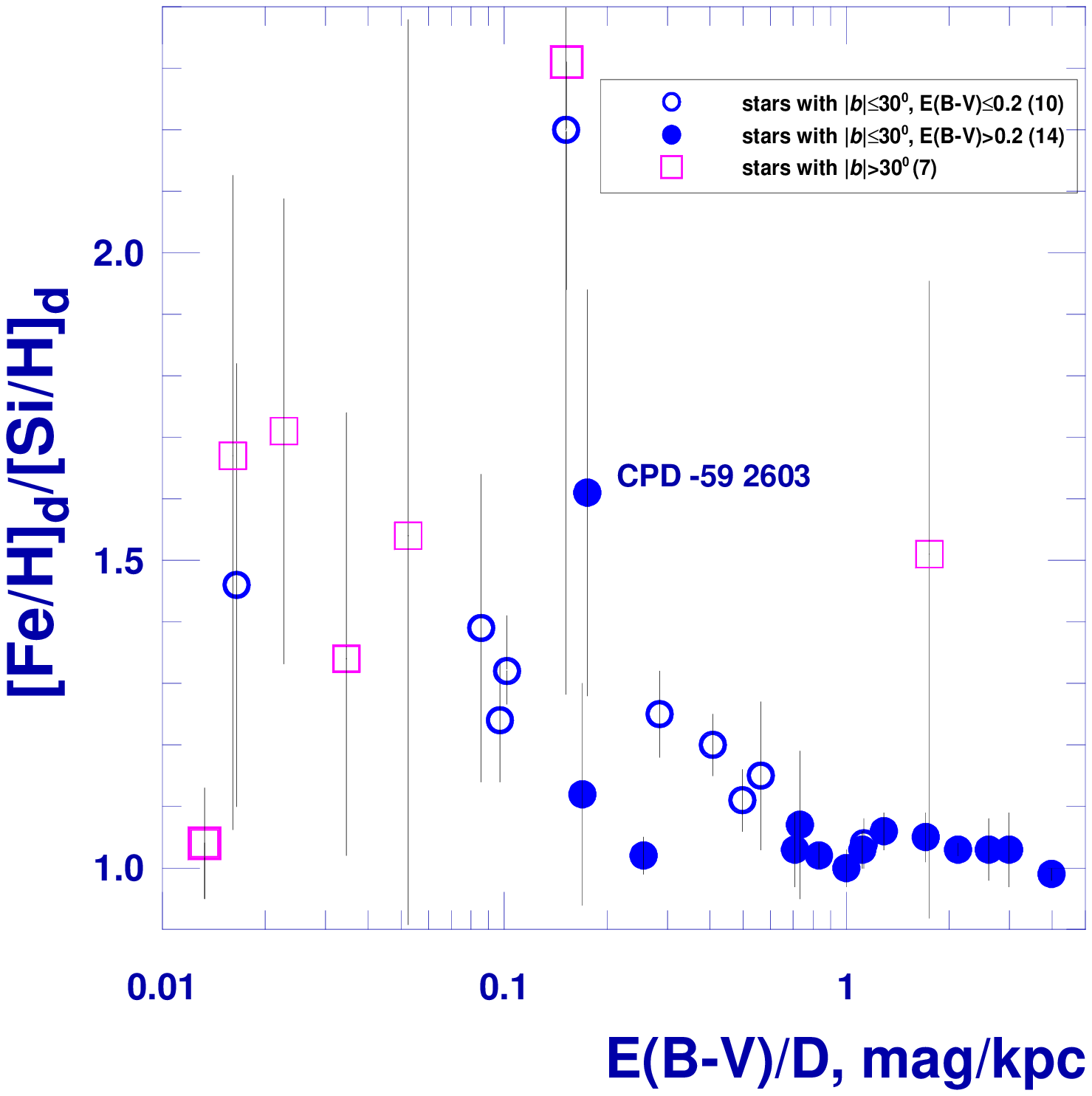}}
\caption{Ratio of dust phase abundances of
(Fe+Mg)/Si, Mg/Si, and Fe/Si with 1$\sigma$ error
bars in dependence on average reddening  for
31 sightlines with joint measurements of three elements.
Open and filled circles show data for sightlines
with diffuse and translucent interstellar clouds, respectively.
Squares correspond to {\rm high-latitude} stars.
}
\label{3m}\ec
\end{figure}

Figure~\ref{3m} shows the ratios as a function of average reddening
for 31 sightlines where joint measurements of abundances of Mg, Si, and Fe
are available.
As follows from Fig.~\ref{3m} (upper panel),
the ratio  (Fe+Mg)/Si$>2$ for all stars.
This ratio  may exceed 3 for low-reddened and {\rm high-latitude} stars.
For high reddened stars (excluding CPD\,--59\,2603),
the ratio  (Fe+Mg)/Si lies in a narrow range from 2.15 to 2.35.
The average value for 13 sightlines is
$\langle {\rm (Fe+Mg)/Si} \rangle =2.25 \pm 0.14$.

The middle and low panels of Fig.~\ref{3m} demonstrate that
magnesium and iron are incorporated into dust particles in unequal parts
for high and low reddened stars and {\rm high-latitude} stars.
For disk stars with $E({\rm B-V}) > 0.20$,
the composition of grains averaged over 13 targets is
Mg$_{1.22}$Fe$_{1.04}$SiO$_{z1}$.
This composition cannot be reproduced by the mixture of
olivine and pyroxene silicates alone with any value of $x$ and $y$.
It indicates that some amount of magnesium or iron (or both)
should be embedded in another population of dust grains,
probably, metal oxides.

When we consider {\rm high-latitude} stars,  averaging over 6 targets
(excluding HD~38666) gives the `grain composition'
Mg$_{1.50}$Fe$_{1.47}$SiO$_{z2}$\footnote{The
values of $z1$, $z2$ are not the same and must lie between 3 and 4 for
a mixture of pyroxene and olivine grains.};
i.e., the ratio Fe/Mg is greater for {\rm high-latitude} stars
than for disk stars.
{\rm This suggests that the destruction of Mg-rich grains
in the warm medium is more effective than of Fe-rich grains.}

\subsection{... + problematic O}

\begin{figure}
\bc
\resizebox{\hsize}{!}{\includegraphics{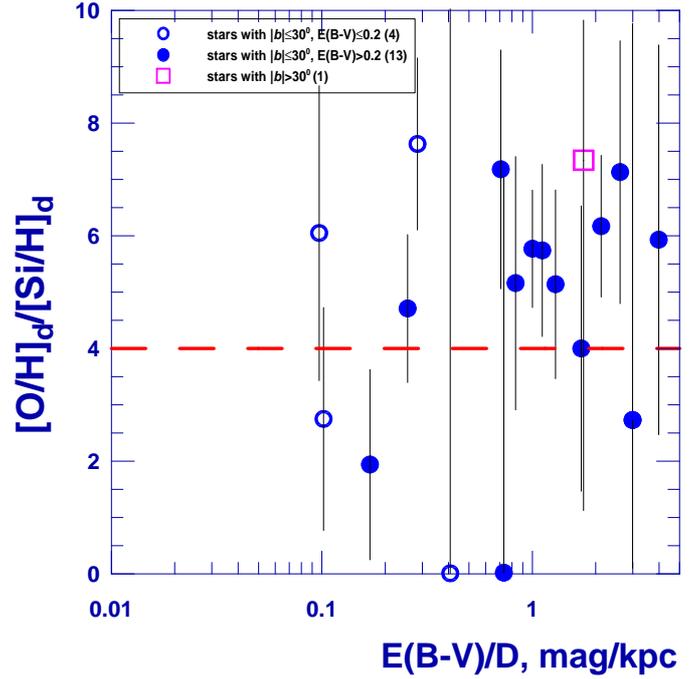}}
\caption{Ratio of dust phase abundances of O and Si
with 1$\sigma$ error bars reduced twice
in dependence on average reddening $E({\rm B-V})/D$.
Open and filled circles show data for disk stars
seen through diffuse and translucent interstellar clouds, respectively.
Squares correspond to sightlines with $|b| > 30\degr$.}
\label{o-si}\ec
\end{figure}

\begin{figure}[htb]
\bc
\resizebox{\hsize}{!}{\includegraphics{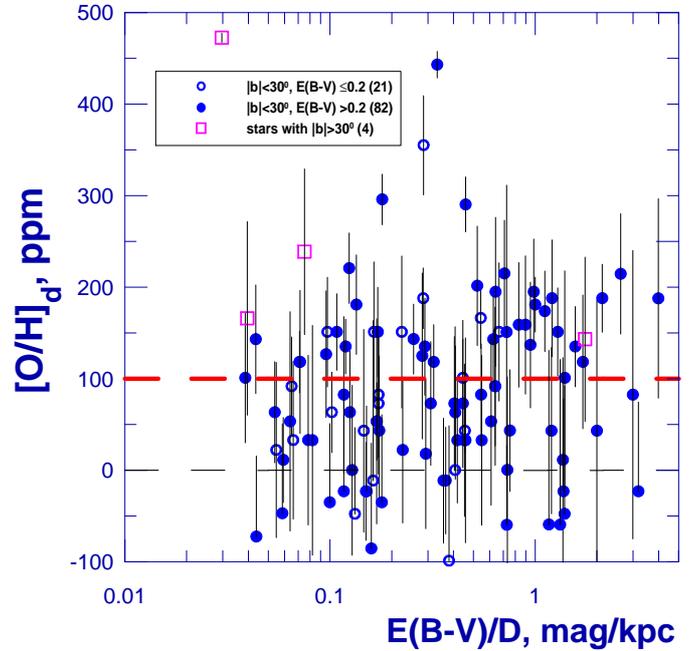}}
\caption{Dust phase abundances of O in ppm
with 1$\sigma$ error bars reduced in twice
in dependence on average reddening $E({\rm B-V})/D$.
Open and filled circles show data for disk stars
seen through diffuse and translucent interstellar clouds, respectively.
Squares correspond to sightlines with $|b| > 30\degr$.}
\label{o-ed1}\ec
\end{figure}

The sample of stars  discussed in Sect.~\ref{ol}
includes 20 sightlines where the abundances of OI have also been measured.
Thus, we can compare abundances of four elements.
We plot oxygen abundances in Figs.~\ref{o-si} and \ref{o-ed1}.
The error bars are twice reduced in comparison with those observed
in these figures.
Unfortunately, the uncertainties  in determining of the gas phase
oxygen abundances are too large to allow any definitive
answer about the trends in oxygen depletion.

\afterpage{\clearpage}

Figure~\ref{3m} clearly shows the excess of both iron and magnesium over
silicon, so if we assume that all Si atoms are incorporated
into olivine, the O to Si ratio must be equal to or exceed 4. This ratio
is plotted in Fig.~\ref{o-si} for 18 sightlines\footnote{Two sightlines
(towards HD~38666 and HD~141637) with $[{\rm O}/{\rm H}]_{\rm d} < 0$
are omitted.}.
It is interesting that, for 10 of 13 stars observed through translucent
clouds, the ratio $[{\rm O}/{\rm H}]_{\rm d}/[{\rm Si}/{\rm H}]_{\rm d} \geq 4$.
This means that we have enough O for silicate particles and, perhaps,
some additional oxygen for producing oxides and ices.
However, this effect may be related to the regional variations in
oxygen abundances as all these stars are located in the bottom part of
Fig.~\ref{all-pol}, i.e., at the galactic longitudes
$l \approx 180\degr - 360\degr$.

In Fig.~\ref{o-ed1} we plot the oxygen dust phase abundances for our
sample of stars. These abundances were found as the difference between
solar oxygen abundance (490~ppm) and observed gas phase abundances.
The data are shown for 107 sightlines (12 sightlines where
[O/H]$_{\rm d} < -150$\,ppm were omitted).
The absence of  trends and correlations as seen for Mg, Si, or Fe
(Fig.~\ref{mgd}) is obvious.
For many directions the dust phase abundances of O are negative and
primarily  related to large observational errors.
We can make a tentative inference about the deficit
(not excess!) of oxygen in the dust phase opposite to the
conclusion of J09. In the case of using  proto-Sun oxygen abundance
([O/H]$_{\sun}= 575$~ppm from Lodders, \cite{lod03}),
things will not get much better,
so it seems too early to search for the `missing oxygen'
in the dust phase (Whittet,~\cite{whi09}).

\begin{figure}
\bc
\resizebox{7.8cm}{!}{\includegraphics{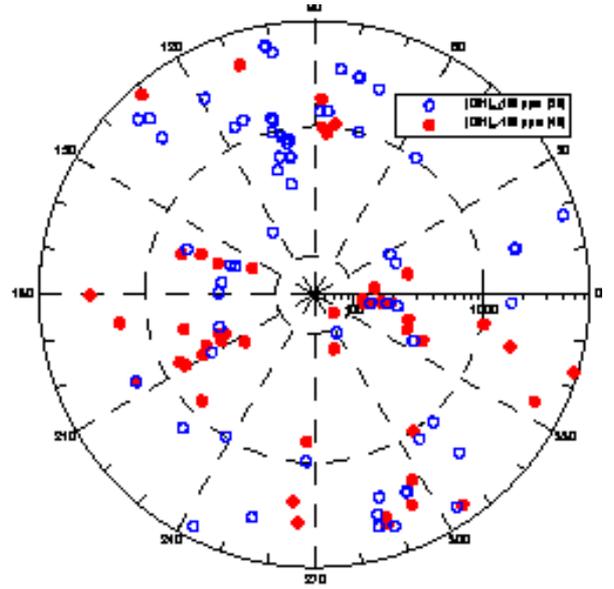}}
\caption{Distance distribution of stars with
[O/H]$_{\rm d} \leq 100$\,ppm (open circles)
and [O/H]$_{\rm d} > 100$\,ppm (filled circles)
in polar representation. The Galactic centre is to the right.
The distance from the Sun is plotted in logarithmic scale.
Number of stars is indicated in parentheses.}
\label{o-pol}\ec
\end{figure}

We divided the stars into two groups: with
[O/H]$_{\rm d} > 100$\,ppm and [O/H]$_{\rm d} \leq 100$\,ppm.
This border value  is calculated as the product $4 \times 25$,
assuming the mean value for Si from Table~\ref{tab1}
and assuming that all Si atoms are tied up into olivine-type
silicates\footnote{Models of dust evolution
predict a dust phase abundance of oxygen at a level of about 130~ppm
at the modern time (see, e.g., Fig.~16 in Zhukovska et al.~\cite{zgt08}).}.
Such a division reveals an interesting galactic distribution of
the `O-rich' and `O-poor' sightlines shown in Fig.~\ref{o-pol}.
It can be seen that the open and filled circles are not well mixed.
There are areas on the sky where the symbols of one type concentrate.
Particularly striking is the zone between
$l \approx 70\degr$ and $l \approx 140\degr$ (Cygnus, Cassiopea, Perseus)
where 23 of 59 stars with reduced O abundance in dust are located.
Perhaps, this behaviour does not merely reflect the observational errors
and indicate the existence of an another non-solar cosmic standard
with enhanced metal abundances in this area.
This hypothesis is partially supported by a higher fraction of metal-rich
Cepheids found at these galactic longitudes (see Fig.~5 in
Pedicelli et al. \cite{petal09}).

\subsection{Correlation with extinction and regional variations}

The elements of {\rm C, O, together with}
Mg, Si, and Fe as studied here contribute to most of the mass of the
interstellar dust. Therefore, it is interesting to study a dependence of
dust phase abundances on the ratio $R_{\rm V}$.
The ratio of total to selective extinction  $R_{\rm V}$
characterizes the visual extinction produced
by dust particles with radii $r \ga 0.05\,\mu$m (e.g., Voshchinnikov,~\cite{v04}).
The search for a correlation between depletions
$D_{\rm O,\,Mg,\,Fe}$ and $R_{\rm V}$ has been attempted by Jensen~(\cite{jens07}), who
finds no correlation for Mg and Fe and a slight trend to increasing OI
depletion with increasing $R_{\rm V}$.

In our sample there are 164 stars with known or calculated
values of $R_{\rm V}$.  We found no significant correlation of dust phase
abundances with  $R_{\rm V}$ either for the total sample or for the disk
stars with low and high reddening. Very probably, the absence of correlation
is a consequence of the mixture of short and long sightlines in different
galactic directions.

We investigated the regional variations of depletions and
$R_{\rm V}$. The difference in iron depletion was found
by Savage \& Bohlin~(\cite{sb79}) for sightlines in Cygnus and
Scorpius--Ophiuchus,
while  Patriarchi et al.~(\cite{pmp03}) and Wegner~(\cite{ww03})
discovered the difference in $R_{\rm V}$ between stars in Cygnus and Carina
and stars belonging to separate associations, respectively.

\begin{figure}
\bc
\resizebox{\hsize}{!}{\includegraphics{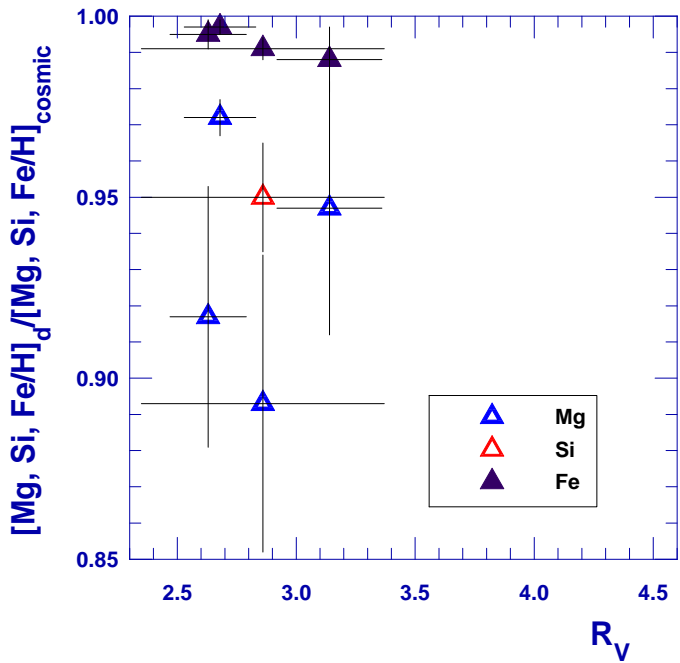}}
\caption{Relative dust phase abundances of Mg, Si, and Fe with 1$\sigma$
error bars in dependence on  $R_{\rm V}$ for stars located in
Perseus. The data correspond to the stars (from left to right):
HD~24398, HD~27778, HD~24912, and HD~23180.
The values of $R_{\rm V}$ were taken from Fitzpatrick \& Massa~(\cite{fm07})
for HD~23180 and HD~27778, Wegner~(\cite{ww03}) for HD~24398  and
Valencic et al.~(\cite{vcg04}) for HD~24912.}
\label{rv-1}\ec
\end{figure}

\begin{figure}
\bc
\resizebox{\hsize}{!}{\includegraphics{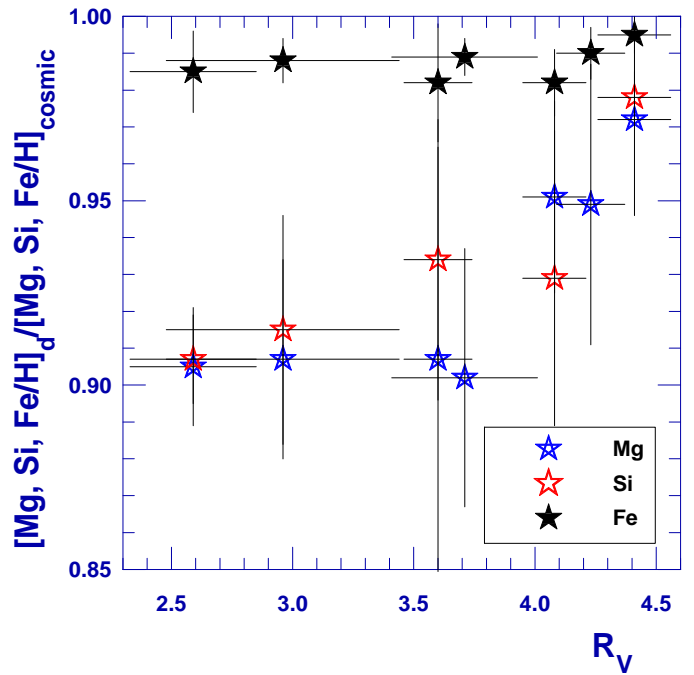}}
\caption{The same as in Fig.~\ref{rv-1} but now
for stars located in Scorpius--Ophiuchus.
The data correspond to the stars (from left to right):
HD~144217, HD~143275, HD~147165, HD~144470, HD~147888, HD~148184, and HD~147933.
The values of $R_{\rm V}$ were taken from Fitzpatrick \& Massa~(\cite{fm07})
for HD~144470,  HD~147165,  HD~147888, and HD~147933 and
Lewis et al.~(\cite{lew09})  for HD~143275, HD~144217, and HD~148184.}
\label{rv-2}\ec
\end{figure}

To exclude the distance effects discussed in Sect.~\ref{sel}
we only considered reddened stars with $D \la 450$\,pc.
From this subsample of 25 sightlines, we separated two groups of stars
more or less closely located on the sky (see also
Figs.~\ref{all-lb} and \ref{all-pol} {\rm and Table~\ref{t00}): four stars in Perseus
($b=-13\degr \div -17\degr$, $l=160\degr - 173\degr$, $D\approx 220 - 420$\,pc,
$N({\rm H})= (1.26 - 1.95) \times 10^{21}\,{\rm cm}^{-2}$,
$E({\rm B-V})=0.27 -0.36$)
and seven stars in Scorpius--Ophiuchus
($b=17\degr - 23\degr$, $l=350\degr - 358\degr$, $D\approx 120 - 200$\,pc,
$N({\rm H})= (1.38 - 5.50) \times 10^{21}\,{\rm cm}^{-2}$,
$E({\rm B-V})=0.21 -0.51$).}
Three of them (HD~147165, HD~147888, and HD~147933) belong to the Eastern
Group of the $\rho$~Oph cloud, while four other stars
belong to the Northern Group (Snow et al.,~\cite{sdw08}).
The measured abundances of Mg, Si, and Fe are compared with
$R_{\rm V}$ in Figs.~\ref{rv-1} and \ref{rv-2}.
For stars in Perseus, the value of $R_{\rm V}$ is lower, on average,
than the mean galactic value $R_{\rm V}=3.1$. In this case the
correlation between  dust phase abundances and $R_{\rm V}$ is absent
(Fig.~\ref{rv-1}). For  stars in Scorpius--Ophiuchus, $R_{\rm V}$
varies from $\sim 2.6$ to $\sim 4.4$. Figure~\ref{rv-2} shows
a clear growth of silicon and magnesium abundances in dust with
increasing $R_{\rm V}$\footnote{\rm A similar trend is seen if we
replace $R_{\rm V}$ by $N({\rm H})$.}.
The Pearson correlation coefficients are
$r_{\rm corr.}= 0.81$ for Mg and $r_{\rm corr.}= 0.86$ for Si.
This is the first evidence of a correlation of dust phase abundances
with the ratio $R_{\rm V}$.
{\rm Although the correlation coefficients are relatively large,
one should keep in mind the small number of sightlines.
Since $R_{\rm V}$ is considered as a measure of grain size
(Whittet,~\cite{w03}), we  conclude that
{\it accretion of Si and Mg atoms on large grains takes place.}}
Note also that obviously the abundance of iron is independent of
the value of $R_{\rm V}$.

\begin{figure}
\bc\resizebox{\hsize}{!}{\includegraphics{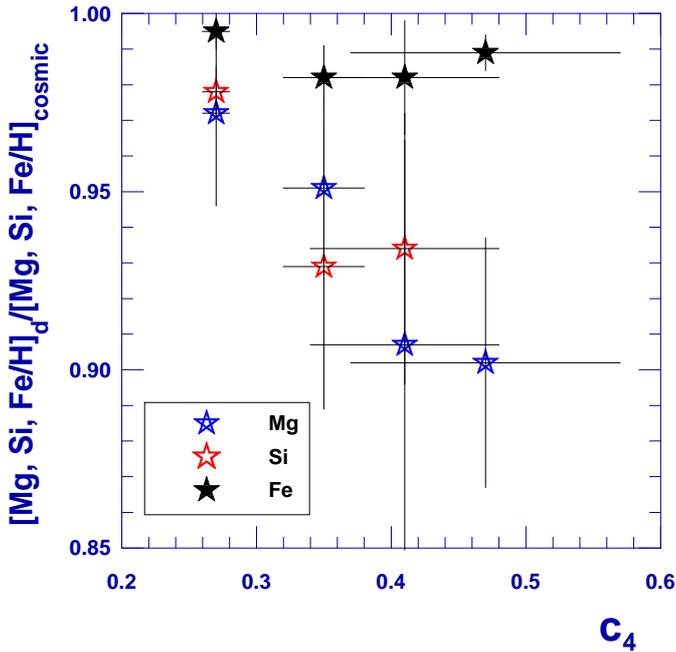}}
\caption{Relative dust phase abundances of Mg, Si, and Fe with 1$\sigma$
error bars as a function of UV extinction curve parameter  $c_4$
{\rm (the strength of the far-UV curvature)} for stars located in Scorpius--Ophiuchus.
The data correspond to the stars (from left to right):
HD~147933, HD~147888, HD~147165, and HD~144470.}
\label{uv2}\ec
\end{figure}

\begin{figure}
\bc
\resizebox{\hsize}{!}{\includegraphics{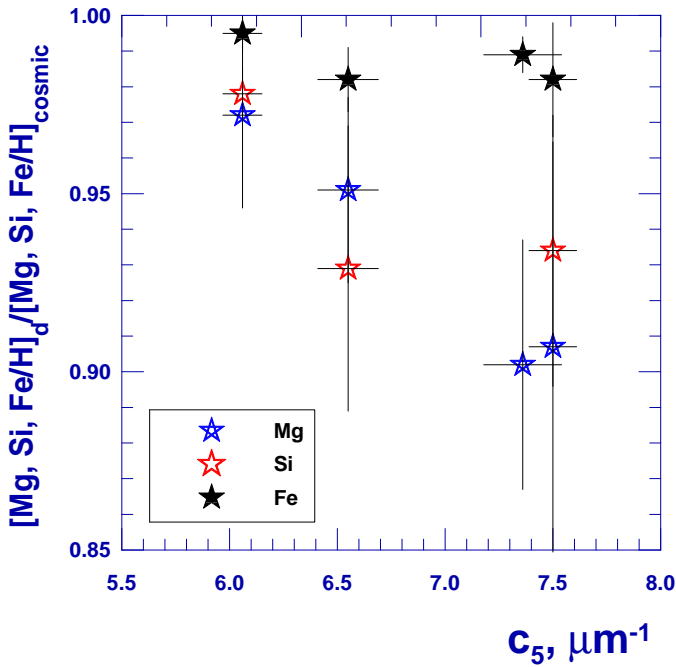}}
\caption{\rm Relative dust phase abundances of Mg, Si, and Fe with 1$\sigma$
error bars as a function of UV extinction curve parameter  $c_5$
(wavenumber from which the far-UV extinction starts to grow)
for stars located in Scorpius--Ophiuchus.
The data correspond to the stars (from left to right):
HD~147933, HD~147888, HD~144470, and HD~147165.}
\label{uv22}\ec
\end{figure}

For three stars in Perseus and four stars in Scorpius--Ophiuchus
the extinction curve in the far-UV is also known.
{\rm Fitzpatrick \& Massa~(\cite{fm07}) fit the observed far-UV extinction
using two parameters $c_4$ and $c_5$, entering into a formula for
the entire UV extinction. These parameters characterize the departure,
in the far-UV, from the extrapolated bump--plus--linear components
and indicate the strength of the far-UV curvature ($c_4$) and
the wavenumber in $\mkm^{-1}$ from which
the far-UV extinction starts to grow ($c_5$).
We interpret parameter $c_4$ as a measure of the relative amount
of small grains (the slope of the size distribution
function, e.g., the power index $q$ in the power-law size distribution
$n(r) \propto r^{-q}$) and parameter
$c_5$ as a measure of the minimum particle size $r_{\min}$
in the dust ensemble.
Unfortunately, the limited data for Perseus do not allow a careful analysis.
For stars located in Scorpius--Ophiuchus the normalized
far-UV extinction is lower than the average Galactic extinction curve,
which points to a deficit of the particles of small sizes.
Figures~\ref{uv2} and \ref{uv22} show the
anticorrelation of the dust phase abundances of Mg and Si with
parameters $c_4$ and $c_5$ (the correlation coefficients are
from $r_{\rm corr.}= -0.70$ to $r_{\rm corr.}= -0.99$).
The shift from right to left in these figures indicates the decrease in $q$
and increase in $r_{\min}$
(see, e.g., Voshchinnikov \& Il'in,~\cite{vi93}; and Fig.~24 in
Voshchinnikov,~\cite{v04}).
This flattening of the size distribution function and growth of minimum
grain size is accompanied by the increase in the dust phase fraction of Mg and
Si; i.e., {\it smaller grains are built up due to accretion of atoms from the
gas}. Apparently, this occurs in clouds with the hydrogen column density
$N({\rm H}) \ga 2 \times 10^{21}\,{\rm cm}^{-2}$
resulting in after the propagation of a low-velocity shock
(Meyers et al.~\cite{msfb85}).
Our arguments in favour of grain growth by accretion
are supported by the results of a detailed interpretation
of extinction for two stars made by Das et al.~(\cite{dvi10}).
They find that for silicate grains
$r_{\min}=0.07\,\mkm$, $q=2.0$ for HD~147933 and
$r_{\min}=0.04\,\mkm$, $q=2.2$ for HD~147165.

The uniform variations in abundances of Mg and Si
with $R_{\rm V}$, $N({\rm H})$, $c_4$, and $c_5$ cannot be explained
by grain coagulation because in this case the gas and dust phase
abundances of elements are kept constant.
}

\section{Conclusions}

We investigated the differences in interstellar dust phase abundances
of Mg, Si, and Fe entering into the composition of silicate and oxide
grains. The distinctions in abundances can be separated into the following
groups.

\begin{enumerate}

\item
A sharp   
distinction in abundances is observed
for sightlines located at low ($|b|<30\degr$) and high ($|b|>30\degr$)
galactic latitudes. This is well known from previous studies.
For {\rm high-latitude} stars the ratios Mg/Si and Fe/Si in dust are close to 1.5.
For disk stars these ratios are reduced to $\sim 1.2$ and $\sim 1.05$
for Mg and Fe, respectively. The derived numbers indicate that
the dust grains  cannot be just a mixture of only
olivine and pyroxene silicates.
Some amount of magnesium or iron (or both) should be embedded into
another population, probably  oxides.

\item
We reveal a clear distinction in abundances for
nearby ($D\la 400$\,pc) and distant ($D\ga 400$\,pc) stars:
decrease in dust phase abundances and correspondingly increase
in gas phase abundances with growth in $D$. We attribute this
distinction to {\rm an observational selection effect:}
a systematic trend  toward lower  observed hydrogen
{\rm column} density for distant stars.
As a result, we obtain a clear trend for abundances as a function of
{\rm any} parameter related to the gas or dust density or distance.

\item
The less pronounced   
difference in abundances
is found for disk stars with low ($E({\rm B-V}) \la 0.2$) and
high ($E({\rm B-V}) \ga 0.2$) reddenings. This reflects the distinction
between sightlines passing through diffuse and translucent interstellar
clouds.

\item
Regional variations of abundances of Mg, Si, and Fe are not evident.
However, for Scorpius--Ophiuchus, we
established an {\rm uniform}
increase of dust phase abundances in Mg and Si with
an increase in the ratio of total to selective extinction  $R_{\rm V}$
and an decrease in {\rm the strength of the far-UV extinction.}
Thus it is valid to say that there is a growth in Mg-Si grains
{\rm due to accretion.}

\end{enumerate}

The uncertainties  in determing of the oxygen abundances are
large and do not allow one to make definitive conclusions
about the oxygen depletion. We can only indicate a
possible regional peculiarity in the zone between
$l \approx 70\degr$ and $l \approx 140\degr$ (Cygnus, Cassiopea, Perseus)
where many stars with reduced O abundance in dust are located.

\acknowledgements{
We thank  Adam Jensen for sending data for silicon,
Andrei Berdyugin for sending the polarimetric results, and
Vladimir Il'in, Jacek Kre{\l}owski,  Piotr Gnaci\'nski, Ralf Siebenmorgen,
{\rm Svitlana Zhukovska}, and Ted Snow
for stimulating discussions. Special thanks go to the anonymous referee
for helpful comments and suggestions.
The work was partly supported by grants
RFBR 07-02-00831, RFBR 10-02-00593a, NTP 2.1.1/665 and NSh 1318.2008.2.}


\clearpage
\newpage
\setcounter{table}{0}
\begin{table*}
\section*{Appendix}
\caption[]{Dust phase abundances of Mg, Si, and Fe in ppm.}    \label{t00}
\bc\begin{tabular}{llrrcrcccc}
\hline\hline\noalign{\smallskip}
\multicolumn{1}{c}{$N$} &
\multicolumn{1}{c}{Star} &
\multicolumn{1}{c}{$l$} &
\multicolumn{1}{c}{$b$} &
\multicolumn{1}{c}{Spectrum} &
\multicolumn{1}{c}{$D$, pc} &
\multicolumn{1}{c}{$E({\rm B-V})$} &
\multicolumn{1}{c}{[Mg/H]$_{\rm d}$} &
\multicolumn{1}{c}{[Si/H]$_{\rm d}$} &
\multicolumn{1}{c}{[Fe/H]$_{\rm d}$} \\
\multicolumn{1}{c}{(1)} &
\multicolumn{1}{c}{(2)} &
\multicolumn{1}{c}{(3)} &
\multicolumn{1}{c}{(4)} &
\multicolumn{1}{c}{(5)} &
\multicolumn{1}{c}{(6)} &
\multicolumn{1}{c}{(7)} &
\multicolumn{1}{c}{(8)} &
\multicolumn{1}{c}{(9)} &
\multicolumn{1}{c}{(10)} \\
\noalign{\smallskip}\hline\noalign{\smallskip}
~~1 &  HD 1383      & 119.02 & --0.89 &  B1II          &   2702   & 0.47 &   32.56 $\pm$ 3.23 &       $\cdots$    &       $\cdots$      \\
~~2 &  HD 5394      & 123.58 & --2.15 &  B0IVpe        &    188   & 0.12 &   34.18 $\pm$ 1.98 &       $\cdots$    &  30.71 $\pm$ 0.44   \\
~~3 &  HD 12323     & 132.91 & --5.87 &  ON9V          &   3586   & 0.21 &   34.18 $\pm$ 1.90 &       $\cdots$    &  30.86 $\pm$ 0.29   \\
~~4 &  HD 13268     & 133.96 & --4.99 &  O8V           &   2391   & 0.36 &   33.19 $\pm$ 2.23 &       $\cdots$    &       $\cdots$      \\
~~5 &  HD 13745     & 134.58 & --4.96 &  O9.7IIn       &   1900   & 0.37 &   33.34 $\pm$ 2.44 &       $\cdots$    &  30.15 $\pm$ 0.62   \\
~~6 &  HD 14434     & 135.08 & --3.82 &  O6.5V         &   4108   & 0.48 &   33.49 $\pm$ 2.13 &       $\cdots$    &       $\cdots$      \\
~~7 &  HD 15137     & 137.46 &  +7.58 &  O9.5II-IIIn   &   3300   & 0.31 &   31.29 $\pm$ 3.37 &       $\cdots$    &  30.12 $\pm$ 0.86   \\
~~8 &  HD 18100     & 217.93 &--62.73 &  B5II-III      &   3100   & 0.05 &   29.08 $\pm$ 4.67 &  16.91 $\pm$ 6.75 &  28.29 $\pm$ 1.53   \\
~~9 &  HD 21856     & 156.32 &--16.75 &  B1V           &    500   & 0.19 &        $\cdots$    &  23.49 $\pm$ 6.51 &  30.69 $\pm$ 0.72   \\
 10 &  HD 22586     & 264.19 &--50.36 &  B2III         &   2020   & 0.06 &   36.92 $\pm$ 1.14 &       $\cdots$    &  30.02 $\pm$ 0.93   \\
 11 &  HD 22928     & 150.28 & --5.77 &  B5III         &    160   & 0.05 &   39.18 $\pm$ 0.23 &       $\cdots$    &  31.49 $\pm$ 0.04   \\
 12 &  HD 22951     & 158.92 &--16.70 &  B0.5V         &    320   & 0.19 &   38.14 $\pm$ 1.21 &       $\cdots$    &       $\cdots$      \\
 13 &  HD 23180     & 160.36 &--17.74 &  B1IVSB        &    219   & 0.29 &   37.71 $\pm$ 1.40 &       $\cdots$    &  31.23 $\pm$ 0.30   \\
 14$^*$ &  HD 23478     & 160.76 &--17.42 &  B3IV          &    240   & 0.28 &        $\cdots$    &       $\cdots$    &       $\cdots$      \\
 15$^*$ &  HD 24190     & 160.39 &--15.90 &  B2V           &    550   & 0.30 &        $\cdots$    &       $\cdots$    &       $\cdots$      \\
 16 &  HD 24398     & 162.29 &--16.69 &  B1Ib          &    301   & 0.27 &   36.49 $\pm$ 1.42 &       $\cdots$    &  31.43 $\pm$ 0.13   \\
 17 &  HD 24534     & 163.08 &--17.14 &  O9.5pe        &    590   & 0.59 &   38.54 $\pm$ 0.21 &  31.35 $\pm$ 0.97 &  31.43 $\pm$ 0.08   \\
 18 &  HD 24760     & 157.35 &--10.09 &  B0.5IV        &    165   & 0.11 &   35.53 $\pm$ 1.85 &       $\cdots$    &  30.95 $\pm$ 0.35   \\
 19 &  HD 24912     & 160.37 &--13.10 &  O7V           &    421   & 0.35 &   35.53 $\pm$ 1.64 &  30.78 $\pm$ 0.49 &  31.33 $\pm$ 0.10   \\
 20 &  HD 27778     & 172.76 &--17.39 &  B3V           &    262   & 0.36 &   38.70 $\pm$ 0.20 &       $\cdots$    &  31.49 $\pm$ 0.06   \\
 21 &  HD 28497     & 208.78 &--37.40 &  B2Vne         &    483   & 0.05 &   32.80 $\pm$ 3.15 &  23.34 $\pm$ 2.13 &       $\cdots$      \\
 22 &  HD 30614     & 144.07 &  14.04 &  O9.5Ia        &    963   & 0.29 &   33.04 $\pm$ 3.04 &       $\cdots$    &  30.55 $\pm$ 0.67   \\
 23 &  HD 34029     & 162.59 &  +4.57 &  G8III+G0III   &     13   & 0.01 &   36.08 $\pm$ 0.79 &  23.28 $\pm$ 2.60 &  29.94 $\pm$ 0.35   \\
 24 &  HD 34816     & 214.83 &--26.24 &  B0.5IV        &    260   & 0.05 &   32.56 $\pm$ 3.34 &       $\cdots$    &       $\cdots$      \\
 25 &  HD 34989     & 194.62 &--15.61 &  B1V           &    490   & 0.10 &        $\cdots$    &       $\cdots$    &  31.13 $\pm$ 0.38   \\
 26 &  HD 35149     & 199.16 &--17.86 &  B1V           &    295   & 0.12 &   29.57 $\pm$ 4.88 &  25.94 $\pm$ 0.97 &  31.04 $\pm$ 0.25   \\
 27 &  HD 35715     & 200.09 &--17.22 &  B2IV          &    370   & 0.04 &        $\cdots$    &       $\cdots$    &  31.08 $\pm$ 0.42   \\
 28 &  HD 36486     & 203.86 &--17.74 &  O9.5II        &    281   & 0.08 &   31.86 $\pm$ 1.43 &  24.64 $\pm$ 1.21 &  30.86 $\pm$ 0.18   \\
 29 &  HD 36822     & 195.40 &--12.29 &  B0.5IV-V      &    330   & 0.08 &   34.18 $\pm$ 2.75 &       $\cdots$    &  31.18 $\pm$ 0.24   \\
 30 &  HD 36861     & 195.05 &--12.00 &  O8IIIf        &    550   & 0.09 &   32.04 $\pm$ 3.10 &       $\cdots$    &  30.95 $\pm$ 0.29   \\
 31 &  HD 37021     & 209.01 &--19.38 &  B0V           &    678   & 0.48 &   38.02 $\pm$ 0.75 &  29.95 $\pm$ 1.53 &  30.98 $\pm$ 0.24   \\
 32 &  HD 37043     & 209.52 &--19.58 &  O9III         &    406   & 0.07 &   32.04 $\pm$ 1.81 &       $\cdots$    &  30.45 $\pm$ 0.45   \\
 33 &  HD 37061     & 208.92 &--19.27 &  B0.5 V        &    476   & 0.53 &   38.63 $\pm$ 0.41 &  30.26 $\pm$ 0.78 &  31.09 $\pm$ 0.14   \\
 34 &  HD 37128     & 205.21 &--17.24 &  B0Iab         &    412   & 0.04 &   33.19 $\pm$ 1.83 &  24.99 $\pm$ 1.88 &  31.02 $\pm$ 0.20   \\
 35 &  HD 37367     & 179.04 & --1.03 &  B2 V SB       &    273   & 0.38 &   35.12 $\pm$ 1.94 &       $\cdots$    &       $\cdots$      \\
 36 &  HD 37468     & 206.82 &--17.34 &  O9.5V         &    370   & 0.06 &   29.57 $\pm$ 3.22 &       $\cdots$    &  30.50 $\pm$ 0.70   \\
 37 &  HD 37903     & 206.85 &--16.54 &  B1.5V         &    719   & 0.33 &   38.29 $\pm$ 0.55 &       $\cdots$    &  31.33 $\pm$ 0.14   \\
 38 &  HD 38087     & 207.07 &--16.26 &  B5V           &    315   & 0.31 &        $\cdots$    &       $\cdots$    &  31.14 $\pm$ 0.16   \\
 39 &  HD 38666     & 237.29 &--27.10 &  O9.5V         &    397   & 0.06 &   21.60 $\pm$ 1.72 &  12.45 $\pm$ 1.88 &  28.78 $\pm$ 0.20   \\
 40 &  HD 38771     & 214.51 &--18.50 &  B0Iab         &    221   & 0.12 &   33.04 $\pm$ 1.22 &       $\cdots$    &  31.05 $\pm$ 0.14   \\
 41 &  HD 40111     & 183.97 &  +0.84 &  B0.5II        &    480   & 0.18 &   32.21 $\pm$ 3.29 &       $\cdots$    &  30.81 $\pm$ 0.60   \\
 42 &  HD 40893     & 180.09 &  +4.34 &  B0IV:         &   2632   & 0.45 &   33.63 $\pm$ 1.77 &       $\cdots$    &  31.00 $\pm$ 0.21   \\
 43 &  HD 41117     & 189.69 & --0.86 &  B2Ia          &    909   & 0.40 &        $\cdots$    &       $\cdots$    &  30.88 $\pm$ 0.38   \\
 44 &  HD 41161     & 164.97 & +12.89 &  O8Vn          &   1400   & 0.21 &        $\cdots$    &       $\cdots$    &  30.81 $\pm$ 0.22   \\
 45 &  HD 42087     & 187.75 &  +1.77 &  B2.5Ib        &   1578   & 0.37 &        $\cdots$    &       $\cdots$    &  31.13 $\pm$ 0.08   \\
 46 &  HD 43384     & 187.99 &  +3.53 &  B3Ia          &   1100   & 0.58 &        $\cdots$    &       $\cdots$    &  30.53 $\pm$ 0.65   \\
 47 &  HD 43818     & 188.49 &  +3.87 &  B0II          &   1623   & 0.52 &   33.04 $\pm$ 1.65 &       $\cdots$    &       $\cdots$      \\
 48 &  HD 46056     & 206.34 & --2.25 &  O8V(n)        &   1670   & 0.49 &        $\cdots$    &       $\cdots$    &  30.99 $\pm$ 0.10   \\
 49 &  HD 46202     & 206.31 & --2.00 &  O9V           &   1670   & 0.47 &        $\cdots$    &       $\cdots$    &  31.28 $\pm$ 0.07   \\
 50 &  HD 47839     & 202.94 &  +2.20 &  O7Ve          &    313   & 0.07 &   25.67 $\pm$ 5.38 &  19.81 $\pm$ 3.55 &       $\cdots$      \\
 51 &  HD 48915     & 227.23 & --8.89 &  A1V           &      3&$\!\!$--0.01&35.63 $\pm$ 4.12 &       $\cdots$    &       $\cdots$      \\
 52 &  HD 52266     & 219.13 & --0.68 &  O9IV          &   1735   & 0.26 &   34.55 $\pm$ 1.61 &       $\cdots$    &       $\cdots$      \\
 53 &  HD 53367     & 223.71 & --1.90 &  B0IV:e        &    780   & 0.74 &        $\cdots$    &       $\cdots$    &  31.32 $\pm$ 0.12   \\
 54 &  HD 53975     & 225.68 & --2.32 &  O7.5V         &   1400   & 0.185&   30.68 $\pm$ 2.46 &       $\cdots$    &  31.10 $\pm$ 0.14   \\
\end{tabular}\ec
\end{table*}
\setcounter{table}{0}
\renewcommand{\footnoterule}{}  
\begin{table*}
\caption[]{(Continued.)}
\bc\begin{tabular}{lcrrcrcccc}
\hline\hline\noalign{\smallskip}
\multicolumn{1}{c}{$N$} &
\multicolumn{1}{c}{Star} &
\multicolumn{1}{c}{$l$} &
\multicolumn{1}{c}{$b$} &
\multicolumn{1}{c}{Spectrum} &
\multicolumn{1}{c}{$D$, pc} &
\multicolumn{1}{c}{$E({\rm B-V})$} &
\multicolumn{1}{c}{[Mg/H]$_{\rm d}$} &
\multicolumn{1}{c}{[Si/H]$_{\rm d}$} &
\multicolumn{1}{c}{[Fe/H]$_{\rm d}$} \\
\multicolumn{1}{c}{(1)} &
\multicolumn{1}{c}{(2)} &
\multicolumn{1}{c}{(3)} &
\multicolumn{1}{c}{(4)} &
\multicolumn{1}{c}{(5)} &
\multicolumn{1}{c}{(6)} &
\multicolumn{1}{c}{(7)} &
\multicolumn{1}{c}{(8)} &
\multicolumn{1}{c}{(9)} &
\multicolumn{1}{c}{(10)} \\
\noalign{\smallskip}\hline\noalign{\smallskip}
~~55 &  HD 54662     & 224.17 & --0.78 &  O6.5V         &   1220   & 0.26 &   36.71 $\pm$ 2.66 &       $\cdots$    &  31.28 $\pm$ 0.18   \\
~~56 &  HD 57060     & 237.82 & --5.37 &  O7Iabfp       &   1870   & 0.14 &   29.80 $\pm$ 5.19 &       $\cdots$    &  30.31 $\pm$ 0.63   \\
~~57 &  HD 57061     & 238.18 & --5.54 &  O9II          &    980   & 0.10 &   25.35 $\pm$ 4.91 &  23.07 $\pm$ 1.24 &  30.45 $\pm$ 0.39   \\
~~58 &  HD 62542     & 255.92 & --9.24 &  B5V           &    396   & 0.36 &        $\cdots$    &       $\cdots$    &  29.77 $\pm$ 1.27   \\
~~59 &  HD 63005     & 242.47 & --0.93 &  O6Vf          &   5200   & 0.28 &   34.67 $\pm$ 1.19 &       $\cdots$    &       $\cdots$      \\
~~60 &  HD 64760     & 262.06 &--10.42 &  B0.5Ib        &    510   & 0.08 &   30.25 $\pm$ 3.92 &       $\cdots$    &  30.34 $\pm$ 0.65   \\
~~61 &  HD 65575     & 266.68 &--12.32 &  B3IVp         &    140   & 0.05 &   39.45 $\pm$ 0.00 &       $\cdots$    &  31.47 $\pm$ 0.03   \\
~~62 &  HD 65818     & 263.48 &--10.28 &  B2II/IIIn     &    290   & 0.06 &   31.67 $\pm$ 5.56 &       $\cdots$    &       $\cdots$      \\
~~63 &  HD 66788     & 245.43 &  +2.05 &  O8V           &   4200   & 0.20 &   32.72 $\pm$ 3.36 &       $\cdots$    &  30.84 $\pm$ 0.33   \\
~~64 &  HD 66811     & 255.98 & --4.71 &  O5Ibnf        &    330   & 0.05 &   26.92 $\pm$ 2.49 &  13.78 $\pm$ 9.48 &  30.28 $\pm$ 0.29   \\
~~65 &  HD 68273     & 262.80 & --7.68 &  WC8+O9I       &    350   & 0.03 &   30.68 $\pm$ 4.64 &  21.44 $\pm$ 3.43 &  29.86 $\pm$ 0.57   \\
~~66 &  HD 69106     & 254.52 & --1.33 &  B0.5II        &   3076   & 0.20 &   34.55 $\pm$ 1.09 &       $\cdots$    &  31.15 $\pm$ 0.45   \\
~~67 &  HD 71634     & 273.32 &--11.52 &  B5III         &    400   & 0.13 &   36.98 $\pm$ 1.53 &       $\cdots$    &       $\cdots$      \\
~~68$^*$ &  HD 72754     & 266.83 & --5.82 &  B2Ia:pshe     &    690   & 0.36 &        $\cdots$    &       $\cdots$    &       $\cdots$      \\
~~69 &  HD 73882     & 260.18 &  +0.64 &  O8.5V         &    759   & 0.67 &        $\cdots$    &       $\cdots$    &  31.08 $\pm$ 0.19   \\
~~70 &  HD 74375     & 275.82 &--10.86 &  B1.5III       &    440   & 0.14 &        $\cdots$    &       $\cdots$    &  30.95 $\pm$ 0.54   \\
~~71 &  HD 75309     & 265.86 & --1.90 &  B2Ib/II       &   2924   & 0.28 &   33.77 $\pm$ 2.30 &       $\cdots$    &       $\cdots$      \\
~~72 &  HD 79186     & 267.36 &  +2.25 &  B5Ia          &    980   & 0.40 &   34.79 $\pm$ 1.71 &       $\cdots$    &       $\cdots$      \\
~~73 &  HD 79351     & 277.69 & --7.37 &  B2IV-V        &    140   & 0.10 &   34.05 $\pm$ 0.00 &       $\cdots$    &       $\cdots$      \\
~~74$^*$ &  HD 88115     & 285.32 & --5.53 &  B1.5IIn       &   3654   & 0.20 &        $\cdots$    &       $\cdots$    &       $\cdots$      \\
~~75 &  HD 90087     & 285.16 & --2.13 &  B2/B3III      &   2716   & 0.30 &        $\cdots$    &  29.80 $\pm$ 1.90 &  30.74 $\pm$ 0.15   \\
~~76 &  HD 91316     & 234.89 & +52.77 &  B1Iab         &   1754   & 0.04 &   25.35 $\pm$ 6.66 &  17.95 $\pm$ 4.01 &       $\cdots$      \\
~~77 &  HD 91597     & 286.86 & --2.37 &  B7/B8IV/V     &   6400   & 0.27 &   33.04 $\pm$ 1.83 &       $\cdots$    &  30.37 $\pm$ 0.41   \\
~~78 &  HD 91651     & 286.55 & --1.72 &  O9V:n         &   2964   & 0.28 &   27.21 $\pm$ 2.92 &       $\cdots$    &  29.98 $\pm$ 0.36   \\
~~79 &  HD 91824     & 285.70 &  +0.07 &  O7V((f))      &   2910   & 0.25 &   30.25 $\pm$ 2.22 &       $\cdots$    &       $\cdots$      \\
~~80 &  HD 91983     & 285.88 &  +0.05 &  B1III         &   2910   & 0.29 &   30.89 $\pm$ 3.37 &       $\cdots$    &       $\cdots$      \\
~~81 &  HD 92554     & 287.60 & --2.02 &  O5III         &   6795   & 0.39 &   27.50 $\pm$ 4.53 &       $\cdots$    &  29.90 $\pm$ 0.67   \\
~~82 &  HD 93030     & 289.60 & --4.90 &  B0V           &    140   & 0.04 &   31.29 $\pm$ 3.44 &       $\cdots$    &  30.73 $\pm$ 0.32   \\
~~83 &  HD 93205     & 287.57 & --0.71 &  O3V           &   3187   & 0.38 &   31.48 $\pm$ 1.30 &       $\cdots$    &  30.37 $\pm$ 0.24   \\
~~84 &  HD 93222     & 287.74 & --1.02 &  O7III((f))    &   2201   & 0.33 &   30.03 $\pm$ 1.92 &       $\cdots$    &  30.25 $\pm$ 0.47   \\
~~85 &  HD 93521     & 183.14 & +62.15 &  O9Vp          &   1760   & 0.04 &   25.67 $\pm$ 9.98 &  15.42 $\pm$ 4.36 &  26.35 $\pm$ 2.31   \\
~~86 &  HD 93843     & 228.24 & --0.90 &  O6III         &   2548   & 0.27 &   31.86 $\pm$ 1.87 &       $\cdots$    &  30.25 $\pm$ 0.33   \\
~~87 &  HD 94493     & 289.01 & --1.18 &  B0.5Iab       &   3888   & 0.23 &   30.03 $\pm$ 1.53 &       $\cdots$    &  29.36 $\pm$ 0.58   \\
~~88 &  HD 99857     & 294.78 & --4.94 &  B1Ib          &   3058   & 0.33 &   32.72 $\pm$ 2.67 &       $\cdots$    &  30.67 $\pm$ 0.30   \\
~~89 &  HD 99890     & 291.75 &  +4.43 &  B0.5V:        &   3070   & 0.24 &   23.20 $\pm$ 6.72 &       $\cdots$    &  29.51 $\pm$ 0.92   \\
~~90 &  HD 100340    & 258.85 & +61.23 &  B1V           &   3000   & 0.04 &   29.33 $\pm$ 3.95 &  26.78 $\pm$ 1.37 &  27.97 $\pm$ 0.88   \\
~~91 &  HD 103779    & 296.85 & --1.02 &  B0.5II        &   3061   & 0.21 &   30.47 $\pm$ 2.33 &       $\cdots$    &  30.67 $\pm$ 0.30   \\
~~92 &  HD 104705    & 297.45 & --0.34 &  B0.5III       &   2082   & 0.28 &   29.57 $\pm$ 2.55 &       $\cdots$    &  29.94 $\pm$ 0.54   \\
~~93 &  HD 106490    & 298.23 &  +3.79 &  B2IV          &    110   & 0.06 &   30.25 $\pm$ 1.31 &       $\cdots$    &  29.90 $\pm$ 0.23   \\
~~94 &  HD 108248    & 300.13 & --0.36 &  B0.5IV        &    100   & 0.20 &   22.02 $\pm$ 5.66 &       $\cdots$    &  29.15 $\pm$ 0.94   \\
~~95$^*$ &  HD 108639    & 300.22 &  +1.95 &  B1III         &    110   & 0.35 &        $\cdots$    &       $\cdots$    &       $\cdots$      \\
~~96 &  HD 109399    & 301.71 & --9.88 &  B1Ib          &   1900   & 0.26 &   33.91 $\pm$ 1.91 &       $\cdots$    &  30.50 $\pm$ 0.37   \\
~~97 &  HD 110432    & 301.96 & --0.20 &  B0.5IIIe      &    301   & 0.51 &        $\cdots$    &       $\cdots$    &  31.47 $\pm$ 0.07   \\
~~98 &  HD 111934    & 303.20 &  +2.51 &  B2Ib          &   2525   & 0.51 &   33.49 $\pm$ 2.70 &       $\cdots$    &       $\cdots$      \\
~~99 &  HD 113904    & 304.67 & --2.49 &  WC5+B0Ia      &   2660   & 0.21 &   35.12 $\pm$ 2.93 &       $\cdots$    &       $\cdots$      \\
100$^*$ &  HD 114886    & 305.52 & --0.83 &  O9IIIn        &   1000   & 0.29 &        $\cdots$    &       $\cdots$    &       $\cdots$      \\
101$^*$ &  HD 115071    & 305.76 &  +0.15 &  B0.5V         &   1200   & 0.44 &        $\cdots$    &       $\cdots$    &       $\cdots$      \\
102 &  HD 116658    & 316.11 & +50.84 &  B1III-IV      &     80   & 0.14 &   34.29 $\pm$ 3.11 &  19.52 $\pm$ 8.65 &  29.41 $\pm$ 1.50   \\
103 &  HD 116781    & 307.05 & --0.07 &  B0IIIe        &   1492   & 0.34 &   31.86 $\pm$ 2.51 &       $\cdots$    &  30.53 $\pm$ 0.48   \\
104 &  HD 116852    & 304.88 &--16.13 &  O9III         &   4832   & 0.21 &   32.04 $\pm$ 2.65 &       $\cdots$    &  29.94 $\pm$ 0.48   \\
105 &  HD 119608    & 320.35 & +43.13 &  B1Ib          &   4200   & 0.12 &   29.08 $\pm$ 2.75 &       $\cdots$    &       $\cdots$      \\
106 &  HD 121263    & 314.07 & +14.19 &  B2.5IV        &    120   & 0.05 &   34.67 $\pm$ 0.66 &       $\cdots$    &  30.77 $\pm$ 0.12   \\
107$^*$ &  HD 121968    & 333.97 & +55.84 &  B1V           &   3800   & 0.15 &        $\cdots$    &       $\cdots$    &       $\cdots$      \\
108 &  HD 122879    & 312.26 &  +1.79 &  B0Ia          &   2265   & 0.36 &   32.04 $\pm$ 2.72 &       $\cdots$    &  30.79 $\pm$ 0.36   \\
109 &  HD 124314    & 312.67 & --0.42 &  O6Vnf         &   1100   & 0.46 &   33.34 $\pm$ 1.61 &       $\cdots$    &  31.02 $\pm$ 0.19   \\
110 &  HD 127972    & 322.77 & +16.67 &  B1.5Vne       &     90   & 0.11 &   22.02 $\pm$ 2.09 &       $\cdots$    &  28.91 $\pm$ 0.29   \\
\end{tabular}\ec
\end{table*}
\setcounter{table}{0}
\begin{table*}
\caption[]{(Continued.)}
\bc\begin{tabular}{llrrcrcccc}
\hline\hline\noalign{\smallskip}
\multicolumn{1}{c}{$N$} &
\multicolumn{1}{c}{Star} &
\multicolumn{1}{c}{$l$} &
\multicolumn{1}{c}{$b$} &
\multicolumn{1}{c}{Spectrum} &
\multicolumn{1}{c}{$D$, pc} &
\multicolumn{1}{c}{$E({\rm B-V})$} &
\multicolumn{1}{c}{[Mg/H]$_{\rm d}$} &
\multicolumn{1}{c}{[Si/H]$_{\rm d}$} &
\multicolumn{1}{c}{[Fe/H]$_{\rm d}$} \\
\multicolumn{1}{c}{(1)} &
\multicolumn{1}{c}{(2)} &
\multicolumn{1}{c}{(3)} &
\multicolumn{1}{c}{(4)} &
\multicolumn{1}{c}{(5)} &
\multicolumn{1}{c}{(6)} &
\multicolumn{1}{c}{(7)} &
\multicolumn{1}{c}{(8)} &
\multicolumn{1}{c}{(9)} &
\multicolumn{1}{c}{(10)} \\
\noalign{\smallskip}\hline\noalign{\smallskip}
111 &  HD 135591    & 320.13 & --2.64 &  O7.5IIIf      &   1250   & 0.22 &        $\cdots$    &       $\cdots$    &  31.16 $\pm$ 0.41   \\
112 &  HD 136298    & 331.32 & +13.82 &  B1.5IV        &    210   & 0.07 &   22.82 $\pm$ 2.00 &       $\cdots$    &  29.56 $\pm$ 0.32   \\
113$^*$ &  HD 137595    & 336.72 & +18.86 &  B3Vn          &    400   & 0.25 &        $\cdots$    &       $\cdots$    &       $\cdots$      \\
114 &  HD 138690    & 333.19 & +11.89 &  B2IV          &    130   & 0.07 &   36.17 $\pm$ 0.80 &       $\cdots$    &  31.12 $\pm$ 0.06   \\
115 &  HD 141637    & 346.10 & +21.70 &  B2.5Vn        &    160   & 0.18 &   34.67 $\pm$ 2.38 &  30.00 $\pm$ 0.87 &  31.12 $\pm$ 0.43   \\
116 &  HD 143018    & 347.21 & +20.23 &  B1V           &    141   & 0.07 &   36.98 $\pm$ 2.18 &  27.83 $\pm$ 0.97 &  31.01 $\pm$ 0.26   \\
117 &  HD 143118    & 338.77 & +11.01 &  B2.5IV        &    140   & 0.02 &   36.56 $\pm$ 0.72 &       $\cdots$    &  30.15 $\pm$ 0.23   \\
118 &  HD 143275    & 350.10 & +22.49 &  B0.3IVe       &    123   & 0.21 &   36.08 $\pm$ 1.08 &  29.65 $\pm$ 1.01 &  31.21 $\pm$ 0.20   \\
119 &  HD 144217    & 353.19 & +23.60 &  B0.5V         &    163   & 0.21 &   36.00 $\pm$ 0.64 &  29.38 $\pm$ 0.40 &  31.13 $\pm$ 0.35   \\
120 &  HD 144470    & 352.76 & +22.76 &  B1V           &    183   & 0.22 &   35.91 $\pm$ 1.38 &       $\cdots$    &  31.26 $\pm$ 0.17   \\
121$^*$ &  HD 144965    & 339.04 &  +8.42 &  B2Vne         &    290   & 0.35 &        $\cdots$    &       $\cdots$    &       $\cdots$      \\
122 &  HD 147165    & 351.33 & +17.00 &  B1IIISB,V     &    137   & 0.41 &   36.08 $\pm$ 3.03 &  30.26 $\pm$ 1.22 &  31.04 $\pm$ 0.52   \\
123$^*$ &  HD 147683    & 344.86 & +10.09 &  B4V           &    280   & 0.39 &        $\cdots$    &       $\cdots$    &       $\cdots$      \\
124 &  HD 147888    & 353.65 & +17.71 &  B3V:SB        &    195   & 0.51 &   37.85 $\pm$ 1.02 &  30.11 $\pm$ 1.29 &  31.02 $\pm$ 0.28   \\
125 &  HD 147933    & 353.68 & +17.70 &  B1.5V         &    118   & 0.47 &   38.70 $\pm$ 1.03 &  31.68 $\pm$ 0.23 &  31.44 $\pm$ 0.16   \\
126 &  HD 148184    & 357.93 & +20.68 &  B1.5Ve        &    160   & 0.44 &   37.76 $\pm$ 1.50 &       $\cdots$    &  31.28 $\pm$ 0.23   \\
127 &  HD 148594    & 350.93 & +13.94 &  B9:V          &    134   & 0.21 &   37.80 $\pm$ 0.54 &       $\cdots$    &       $\cdots$      \\
128 &  HD 149404    & 340.54 &  +3.01 &  O9Ia          &    908   & 0.62 &        $\cdots$    &       $\cdots$    &  30.98 $\pm$ 0.11   \\
129 &  HD 149757    &   6.28 & +23.59 &  O9.5Vnn       &    146   & 0.31 &   37.56 $\pm$ 0.49 &  30.45 $\pm$ 0.40 &  31.43 $\pm$ 0.04   \\
130 &  HD 149881    &  31.37 & +36.23 &  B0.5III       &   2100   & 0.11 &   24.30 $\pm$ 1.40 &  18.27 $\pm$11.41 &  28.05 $\pm$ 2.47   \\
131 &  HD 151804    & 343.62 &  +1.94 &  O8Iab         &   1254   & 0.30 &   33.34 $\pm$ 3.92 &       $\cdots$    &  30.77 $\pm$ 0.46   \\
132$^*$ &  HD 151805    & 343.20 &  +1.59 &  B1Ib          &   6009   & 0.43 &        $\cdots$    &       $\cdots$    &       $\cdots$      \\
133 &  HD 152236    & 343.03 &  +0.87 &  B1Ia          &    612   & 0.60 &        $\cdots$    &       $\cdots$    &  31.35 $\pm$ 0.14   \\
134 &  HD 152590    & 344.84 &  +1.83 &  O7.5V         &   1800   & 0.46 &   34.43 $\pm$ 2.03 &  30.45 $\pm$ 0.79 &  31.08 $\pm$ 0.10   \\
135 &  HD 154368    & 349.97 &  +3.22 &  O9Ib          &   1046   & 0.76 &        $\cdots$    &       $\cdots$    &  31.08 $\pm$ 0.26   \\
136 &  HD 155806    & 352.59 &  +2.87 &  O7.5Ve        &    860   & 0.28 &   33.04 $\pm$ 3.38 &       $\cdots$    &  30.94 $\pm$ 0.55   \\
137 &  HD 156110    &  70.99 & +35.91 &  B3Vn          &    720   & 0.03 &   30.89 $\pm$ 0.78 &       $\cdots$    &       $\cdots$      \\
138 &  HD 157246    & 334.64 &--11.48 &  B1Ib          &    348   & 0.06 &   32.88 $\pm$ 2.65 &       $\cdots$    &  30.71 $\pm$ 0.44   \\
139 &  HD 157857    &  12.97 & +13.31 &  O7V           &   1902   & 0.43 &   34.18 $\pm$ 2.21 &       $\cdots$    &       $\cdots$      \\
140 &  HD 158926    & 351.74 & --2.21 &  B2IV          &    220   & 0.10 &        $\cdots$    &  25.16 $\pm$ 0.95 &  30.15 $\pm$ 0.22   \\
141 &  HD 160578    & 351.04 & --4.72 &  B1.5III       &    142   & 0.08 &   31.67 $\pm$ 4.37 &  26.90 $\pm$ 2.42 &  30.86 $\pm$ 0.49   \\
142 &  HD 164740    &   5.97 & --1.17 &  O7.5V(n)      &   1330   & 0.86 &        $\cdots$    &       $\cdots$    &  31.48 $\pm$ 0.03   \\
143 &  HD 165024    & 343.33 &--13.82 &  B2Ib          &    250   & 0.05 &   33.91 $\pm$ 2.00 &       $\cdots$    &  30.86 $\pm$ 0.34   \\
144 &  HD 165955    & 357.41 & --7.43 &  B1Vnp         &   1640   & 0.21 &   31.48 $\pm$ 2.25 &       $\cdots$    &       $\cdots$      \\
145 &  HD 167264    &  10.46 & --1.74 &  B0.5Ia        &   1514   & 0.30 &   36.08 $\pm$ 2.83 &       $\cdots$    &       $\cdots$      \\
146 &  HD 167756    & 351.47 &--12.30 &  B0.5Iab?      &   4230   & 0.07 &   31.67 $\pm$ 2.49 &  20.65 $\pm$ 4.78 &  30.05 $\pm$ 0.50   \\
147 &  HD 168076    &  16.94 &  +0.84 &  O5V           &   1820   & 0.76 &        $\cdots$    &       $\cdots$    &  31.08 $\pm$ 0.45   \\
148 &  HD 170740    &  21.06 & --0.53 &  B2V           &    235   & 0.47 &        $\cdots$    &       $\cdots$    &  31.41 $\pm$ 0.12   \\
149 &  HD 175360    &  12.53 &--11.29 &  B6III         &    270   & 0.12 &   35.82 $\pm$ 1.18 &       $\cdots$    &       $\cdots$      \\
150 &  HD 177989    &  17.81 &--11.88 &  B2II          &   5021   & 0.22 &   34.18 $\pm$ 1.62 &       $\cdots$    &  30.86 $\pm$ 0.27   \\
151 &  HD 179406    &  28.23 & --8.31 &  B3IVvar       &    227   & 0.31 &        $\cdots$    &       $\cdots$    &  31.17 $\pm$ 0.33   \\
152 &  HD 184915    &  31.77 &--13.29 &  B0.5IIIne     &    700   & 0.22 &   36.49 $\pm$ 2.44 &       $\cdots$    &       $\cdots$      \\
153 &  HD 185418    &  53.60 & --2.17 &  B0.5 V        &   1027   & 0.47 &   36.25 $\pm$ 1.09 &       $\cdots$    &  31.21 $\pm$ 0.21   \\
154 &  HD 186994    &  78.62 & +10.06 &  B0III         &   2500   & 0.16 &        $\cdots$    &       $\cdots$    &  30.34 $\pm$ 0.42   \\
155 &  HD 188209    &  80.99 & +10.09 &  O9.5Ib        &   2210   & 0.15 &        $\cdots$    &       $\cdots$    &  30.58 $\pm$ 0.71   \\
156 &  HD 190918    &  72.65 &  +2.06 &  WN4+O9.7Iab   &   2290   & 0.41 &   31.09 $\pm$ 2.25 &       $\cdots$    &       $\cdots$      \\
157 &  HD 192035    &  83.33 &  +7.76 &  B0III-IVn     &   2800   & 0.35 &   36.25 $\pm$ 1.11 &       $\cdots$    &       $\cdots$      \\
158 &  HD 192639    &  74.90 &  +1.48 &  O8V           &    999   & 0.61 &   34.43 $\pm$ 1.62 &       $\cdots$    &  31.14 $\pm$ 0.21   \\
159 &  HD 195965    &  85.71 &  +5.00 &  B0V           &   1300   & 0.22 &   34.05 $\pm$ 1.04 &  27.61 $\pm$ 4.47 &  30.87 $\pm$ 0.10   \\
160 &  HD 197512    &  87.89 &  +4.63 &  B1V           &   1614   & 0.29 &        $\cdots$    &       $\cdots$    &  31.18 $\pm$ 0.12   \\
161 &  HD 198478    &  85.75 &  +1.49 &  B3Ia          &    890   & 0.57 &   37.29 $\pm$ 1.32 &       $\cdots$    &       $\cdots$      \\
162 &  HD 198781    &  99.94 & +12.61 &  B2IV          &    768   & 0.31 &   35.73 $\pm$ 1.22 &       $\cdots$    &       $\cdots$      \\
163$^*$ &  HD 199579    &  87.50 & --0.30 &  B0.5V         &    990   & 0.33 &        $\cdots$    &       $\cdots$    &       $\cdots$      \\
164 &  HD 201345    &  78.44 & --9.54 &  O9V           &   2570   & 0.17 &   30.68 $\pm$ 2.75 &       $\cdots$    &       $\cdots$      \\
165 &  HD 202347    &  88.22 & --2.08 &  B1V           &   1300   & 0.19 &   34.90 $\pm$ 2.17 &       $\cdots$    &  30.79 $\pm$ 0.34   \\
166 &  HD 202904    &  80.98 &--10.05 &  B2Vne         &    276   & 0.13 &        $\cdots$    &  27.83 $\pm$ 4.10 &  30.82 $\pm$ 0.76   \\
\end{tabular}\ec
\end{table*}
\setcounter{table}{0}
\renewcommand{\footnoterule}{}  
\begin{table*}
\caption[]{(Continued.)}
\bc\begin{tabular}{llrrcrcccc}
\hline\hline\noalign{\smallskip}
\multicolumn{1}{c}{$N$} &
\multicolumn{1}{c}{Star} &
\multicolumn{1}{c}{$l$} &
\multicolumn{1}{c}{$b$} &
\multicolumn{1}{c}{Spectrum} &
\multicolumn{1}{c}{$D$, pc} &
\multicolumn{1}{c}{$E({\rm B-V})$} &
\multicolumn{1}{c}{[Mg/H]$_{\rm d}$} &
\multicolumn{1}{c}{[Si/H]$_{\rm d}$} &
\multicolumn{1}{c}{[Fe/H]$_{\rm d}$} \\
\multicolumn{1}{c}{(1)} &
\multicolumn{1}{c}{(2)} &
\multicolumn{1}{c}{(3)} &
\multicolumn{1}{c}{(4)} &
\multicolumn{1}{c}{(5)} &
\multicolumn{1}{c}{(6)} &
\multicolumn{1}{c}{(7)} &
\multicolumn{1}{c}{(8)} &
\multicolumn{1}{c}{(9)} &
\multicolumn{1}{c}{(10)} \\
\noalign{\smallskip}\hline\noalign{\smallskip}
167 &  HD 203374    & 100.51 &  +8.62 &  B0IVpe        &    820   & 0.60 &   34.30 $\pm$ 2.01 &  28.93 $\pm$ 3.01 &  30.82 $\pm$ 0.31   \\
168 &  HD 203532    & 309.46 &--31.74 &  B5V           &    211   & 0.28 &   36.17 $\pm$ 3.46 &       $\cdots$    &       $\cdots$      \\
169 &  HD 206267    &  99.29 &  +3.74 &  O6V           &    814   & 0.52 &   36.56 $\pm$ 1.15 &       $\cdots$    &  31.33 $\pm$ 0.13   \\
170 &  HD 206773    &  99.80 &  +3.62 &  B0V           &    597   & 0.45 &   34.30 $\pm$ 1.22 &       $\cdots$    &       $\cdots$      \\
171 &  HD 207198    & 103.14 &  +6.99 &  O9II          &   1216   & 0.54 &   37.29 $\pm$ 0.52 &       $\cdots$    &  31.32 $\pm$ 0.09   \\
172 &  HD 207308    & 103.11 &  +6.82 &  B0.7III-IVn   &   1470   & 0.52 &   36.71 $\pm$ 1.00 &       $\cdots$    &  31.14 $\pm$ 0.19   \\
173 &  HD 207538    & 101.60 &  +4.67 &  O9.5V         &    880   & 0.64 &   36.71 $\pm$ 1.01 &       $\cdots$    &  31.32 $\pm$ 0.19   \\
174 &  HD 208440    & 104.03 &  +6.44 &  B1V           &    620   & 0.34 &   34.18 $\pm$ 1.98 &       $\cdots$    &       $\cdots$      \\
175$^*$ &  HD 208947    & 106.55 &  +9.00 &  B2V           &    500   & 0.19 &        $\cdots$    &       $\cdots$    &       $\cdots$      \\
176 &  HD 209339    & 104.58 &  +5.87 &  B0IV          &    980   & 0.35 &   33.77 $\pm$ 1.34 &       $\cdots$    &  30.92 $\pm$ 0.18   \\
177 &  HD 210121    &  56.88 &--44.46 &  B9V           &    223   & 0.38 &        $\cdots$    &       $\cdots$    &  29.86 $\pm$ 0.88   \\
178 &  HD 210809    &  99.85 & --3.13 &  O9Ib          &   3961   & 0.31 &   31.86 $\pm$ 3.01 &       $\cdots$    &       $\cdots$      \\
179 &  HD 210839    & 103.83 &  +2.61 &  O6Iab         &   1260   & 0.57 &   35.82 $\pm$ 1.09 &       $\cdots$    &  31.13 $\pm$ 0.16   \\
180 &  HD 212791    & 101.64 & --4.30 &  B8            &    370   & 0.06 &   34.30 $\pm$ 3.98 &       $\cdots$    &       $\cdots$      \\
181 &  HD 214680    &  96.65 &--16.98 &  O9V           &    610   & 0.08 &   30.68 $\pm$ 4.09 &       $\cdots$    &  31.22 $\pm$ 0.31   \\
182 &  HD 214993    &  97.65 &--16.18 &  B1.5IIIn      &    610   & 0.10 &   35.01 $\pm$ 2.88 &       $\cdots$    &  30.05 $\pm$ 1.18   \\
183 &  HD 215733    &  85.16 &--36.35 &  B1II          &   2900   & 0.10 &   30.89 $\pm$ 5.87 &  21.93 $\pm$ 5.74 &  29.41 $\pm$ 1.04   \\
184 &  HD 218376    & 109.96 & --0.79 &  B1III         &    383   & 0.23 &   34.79 $\pm$ 4.56 &       $\cdots$    &  30.73 $\pm$ 0.85   \\
185$^*$ &  HD 218915    & 108.06 & --6.89 &  O9.5Iabe      &   3660   & 0.26 &        $\cdots$    &       $\cdots$    &       $\cdots$      \\
186 &  HD 219188    &  83.03 &--50.17 &  B0.5III       &   1064   & 0.08 &   33.34 $\pm$ 6.30 &  27.72 $\pm$ 2.70 &       $\cdots$      \\
187 &  HD 220057    & 112.13 &  +0.21 &  B2IV          &   1421   & 0.24 &   36.56 $\pm$ 1.41 &       $\cdots$    &       $\cdots$      \\
188 &  HD 224151    & 115.44 & --4.64 &  B0.5II-III    &   1355   & 0.42 &   32.88 $\pm$ 1.70 &       $\cdots$    &  30.45 $\pm$ 0.37   \\
189 &  HD 224572    & 115.55 & --6.36 &  B1V           &    340   & 0.19 &   36.00 $\pm$ 2.16 &       $\cdots$    &  31.00 $\pm$ 0.53   \\
190 &  HD 232522    & 130.70 & --6.71 &  B1II          &   5438   & 0.21 &   31.48 $\pm$ 1.94 &       $\cdots$    &       $\cdots$      \\
191 &  HD 303308    & 287.59 & --0.61 &  O3V           &   3631   & 0.45 &   32.56 $\pm$ 2.16 &       $\cdots$    &  30.28 $\pm$ 0.41   \\
192 &  HD 308813    & 294.79 & --1.61 &  O9.5V         &   2398   & 0.28 &   32.56 $\pm$ 2.08 &       $\cdots$    &       $\cdots$      \\
193 &  BD +35 4258  &  77.19 & --4.74 &  B0.5 Vn       &   3093   & 0.25 &   32.72 $\pm$ 2.97 &       $\cdots$    &  30.22 $\pm$ 0.64   \\
194$^*$ &  BD +53 2820  & 101.24 & --1.69 &  B0IV:n        &   4506   & 0.37 &        $\cdots$    &       $\cdots$    &       $\cdots$      \\
195 &  CPD -59 2603 & 287.59 & --0.69 &  O7V           &   2630   & 0.46 &   32.72 $\pm$ 1.56 &  18.91 $\pm$ 3.64 &  30.37 $\pm$ 0.32   \\
196$^*$ &  CPD -69 1743 & 303.71 & --7.35 &  B1Vn          &   4700   & 0.30 &        $\cdots$    &       $\cdots$    &       $\cdots$      \\
\noalign{\smallskip}\hline\noalign{\smallskip}
\multicolumn{3}{@{}l@{}}{\rm Notes.}\\
\multicolumn{10}{@{}l@{}}{Dust phase abundances are calculated as difference
between solar abundances ([Mg/H]$_{\sun}= 39.8$~ppm,
[Si/H]$_{\sun}= 32.4$~ppm, [Fe/H]$_{\sun}= 31.6$} \\
\multicolumn{10}{@{}l@{}}{ppm, Asplund et al.,~\cite{agss09}) and gas phase abundances.} \\
\multicolumn{10}{@{}l@{}}{$^*$ For these stars only gas phase abundances of O are
measured.} \\
\end{tabular}\ec
\end{table*}
\end{document}